\documentclass[final,3p,times]{elsarticle}
 
\usepackage{amssymb}
\usepackage{amsmath}
\usepackage{amsthm}
\usepackage{amsfonts}
\usepackage{mathtools}
\usepackage{dsfont}
\usepackage{bm}
\usepackage{booktabs}
\usepackage{cleveref}
\crefformat{equation}{(#2#1#3)}
\Crefname{algocfline}{Algorithm}{Algorithms}
\Crefname{algocf}{line}{lines}
\Crefname{AlgoLine}{Line}{Lines}
\crefname{AlgoLine}{line}{lines}
\usepackage{xcolor}

\usepackage[ruled,vlined,linesnumbered]{algorithm2e}

\newcommand{\mnorm}[1]{{\left\vert\kern-0.25ex\left\vert\kern-0.25ex\left\vert #1 
    \right\vert\kern-0.25ex\right\vert\kern-0.25ex\right\vert}}

\newtheorem{definition}{Definition} 
\newtheorem{theorem}{Theorem}
\newtheorem{lemma}{Lemma}
\newtheorem{corollary}{Corollary}
\newtheorem{remark}{Remark}

\newcommand{\eg}{{\it e.g.}}
\newcommand{\ie}{{\it i.e.}}

\newcommand{\fix}[1]{#1}

\newcommand{\intSet}{\mathbb{Z}}
\newcommand{\realSet}{\mathbb{R}}

\newcommand{\plSet}{\mathbf{N}}
\newcommand{\chSet}{\mathbf{M}}
\newcommand{\opSet}{\mathcal{O}}

\newcommand{\coordFactor}{\mathbf{w}}
\usepackage{subcaption}

\journal{Transportation Research Part C}

\begin{document}
\begin{frontmatter}

\title{Iterative Negotiation and Oversight:\\ A Case Study in Decentralized Air Traffic Management}


\author{Jaehan Im}
\author{John-Paul Clarke}
\author{Ufuk Topcu}
\author{David Fridovich-Keil}

\address{The University of Texas at Austin, Austin, Texas, United States}

\begin{abstract}
Achieving consensus among \fix{self-interested} agents remains challenging in decentralized multi-agent systems, where agents often have conflicting preferences. 
Existing coordination methods enable agents to reach consensus without a centralized coordinator, but do not provide formal guarantees on system-level objectives such as efficiency or fairness. 
To address this limitation, we propose \fix{a regulated decentralized negotiation framework} that augments a decentralized negotiation mechanism with \fix{limited regulatory} oversight. 
The framework builds upon the trading auction for consensus, enabling \fix{self-interested} agents with conflicting preferences to negotiate through asset trading while \fix{avoiding direct disclosure of private asset valuations}. 
We introduce an oversight mechanism, which implements a taxation-like intervention that guides decentralized negotiation toward system-efficient and equitable outcomes while also regulating how fast the framework converges. 
We establish theoretical guarantees of finite-time termination and derive bounds linking system efficiency and convergence rate to the level of \fix{regulatory intervention}. 
A case study based on the collaborative trajectory options program, a rerouting initiative in U.S. air traffic management, demonstrates that the framework can reliably achieve consensus among \fix{self-interested} airspace sector managers, and reveals how the level of \fix{regulatory intervention} regulates the relationship between system efficiency and convergence speed. 
Taken together, the theoretical and experimental results indicate that the proposed framework provides \fix{a mechanism for regulated decentralized coordination that preserves noncooperative final selection while safeguarding system-level objectives}.
\end{abstract}


\end{frontmatter}


\section{Introduction} \label{section:Introduction}

Multi-agent systems frequently require agents to reach a shared decision, such as in task allocation~\cite{cbba}, formation control~\cite{amirkhani2022consensus}, or scheduling~\cite{shoham2008multiagent}. 
Achieving such agreement becomes particularly challenging when agents behave in a self-interested manner, possess conflicting preferences over multiple choices, and prefer not to disclose \fix{certain private quantities, such as asset valuations or operational preferences} \cite{cbba}. 
While these challenges already make agreement difficult, a compounding issue arises when consensus alone is insufficient to ensure system-level objectives such as efficiency, fairness, or safety.
In the absence of a mechanism that safeguards system-level objectives, decentralized negotiation may converge to equilibria that are stable from each agent's perspective but misaligned with system-level objectives.
This limitation becomes particularly critical in safety-sensitive settings, where poorly coordinated outcomes may be suboptimal or even unsafe depending on the application \cite{taco,cooperative_nego,Nego_dacr}.

Existing decentralized consensus mechanisms enable agents to reach consensus without a central coordinator or substantial information sharing \cite{cbba,amirkhani2022consensus,bertsekas1992auction,li2019survey,marden2012loglinear}. 
However, these methods largely focus on reaching agreement---often via heuristic mechanisms---and do not provide formal guarantees on efficiency, fairness, or safety~\cite{cbba,bertsekas1992auction,marden2012loglinear,coadpt}.
In contrast, methods that can explicitly enforce system-level objectives almost always rely on cooperative behaviors or centralized decision-making~\cite{cooperative_nego,chen2013coopctrl,chen2023fairness}.  
Taken together, the current literature lacks mechanisms that \fix{operate in the intermediate regime between purely decentralized negotiation and centralized enforcement: preserving decentralized negotiation while introducing oversight to safeguard system-level objectives} \cite{taco,coadpt,microsimulation2020}.

We adopt a two-layer perspective in which a decentralized negotiation mechanism aligns self-interested behavior toward consensus, while \fix{a regulatory oversight layer provides system-level performance control when autonomous negotiation alone is insufficient}. 
We build on the trading auction for consensus (\texttt{TACo}) \cite{taco}, an auction-based mechanism that enables \fix{self-interested} agents to reach agreement through the exchange of secondary assets such as monetary tokens or emission credits. 
\texttt{TACo} is particularly suitable as a negotiation mechanism because it relies solely on broadcast communication and requires no \fix{direct} disclosure of private valuations of the traded assets. 
Despite these advantages, \texttt{TACo} alone does not directly optimize system-level objectives. 
We therefore introduce an oversight mechanism that performs two functions: (i) it adaptively adjusts a coordination level between agents based on observed trading behavior, and (ii) it separately influences the trading process through a taxation-like intervention. 
Together, the oversight mechanism \fix{guides candidate generation without replacing decentralized negotiation with centralized assignment, steering} the negotiation toward more efficient and fair outcomes while regulating the convergence rate.

The contributions of this work are threefold. 
(i) We propose an iterative negotiation and oversight framework that
integrates decentralized negotiation with \fix{limited regulatory oversight}, enabling \fix{consensus among self-interested agents while actively managing system-level performance}.
(ii) We provide theoretical guarantees, proving finite-time termination and deriving explicit bounds that connect the taxation parameter to system-level optimality and convergence speed. 
These results highlight the framework’s applicability to \fix{regulated decentralized coordination problems in safety-critical domains}.
(iii) We evaluate the framework using a case study based on the collaborative trajectory options program \cite{faa_ac90115, li2017ctop, cruciol2015ctop}, demonstrating improvements in efficiency and fairness while \fix{preserving decentralized negotiation among self-interested agents}.
\section{Related Works}

\paragraph{Decentralized and Noncooperative Coordination Mechanisms}
Research on decentralized coordination has developed negotiation-based~\cite{amirkhani2022consensus}, auction-based~\cite{cbba, bertsekas1992auction}, and learning-based mechanisms~\cite{li2019survey, marden2012loglinear} that enable agents to reach agreement while preserving autonomy and limiting disclosure of private preferences~\cite{shoham2008multiagent}. 
These methods scale well in distributed settings and have been applied to task allocation and resource management~\cite{decentCoupled, sciencedirect2022decentralizedtask}. 
\fix{Recent work on decentralized noncooperative games further considers coupled system constraints, where self-interested agents optimize private objectives while maintaining global feasibility through decentralized primal-dual updates and local communication~\cite{decentCoupled}.}
Despite their utility, these approaches primarily address agreement, allocation, or equilibrium computation under a fixed problem formulation, rather than regulating system-level behavior. 
\fix{Thus, their guarantees do not directly characterize the regulated consensus setting considered here, where self-interested agents negotiate over a common operational candidate while system-level efficiency, fairness, and safety must be safeguarded.}

\fix{Distributed optimization methods such as the \emph{alternating direction method of multipliers} (ADMM) provide strong convergence guarantees for convex coordination problems by decomposing a global objective into local subproblems with consensus constraints~\cite{admm,parallelAdmm,decompositionMPC,decompositionTutorial}. 
However, they typically assume cooperative agents optimizing a shared objective and require exchange of primal and dual variables, making them less suitable for strictly noncooperative and privacy-sensitive settings.}

\paragraph{System-Level Safeguard-Embedded Coordination}
A complementary line of work embeds safety, equity, and efficiency requirements directly into centralized optimization, contract-based coordination, or global constraint enforcement schemes~\cite{chen2019prescribed,im2024coordination,delahaye2014robust,contractAssembly,chen2023fairness,gurtner2019fairness}. 
\fix{The main advantage of these approaches is that the coordinator can often provide explicit guarantees on constraint satisfaction, worst-case performance, or equitable treatment across agents, typically under convexity, robustness, or enforceability assumptions.}
However, these guarantees usually rely on cooperative agents, shared information, or \fix{a central authority capable of determining or enforcing system-wide decisions}, making them \fix{less directly applicable to} noncooperative and decentralized architectures in which agents act autonomously and may withhold private information. 
\fix{Even when limited decentralization is introduced through decomposition~\cite{admm,decompositionMPC,decompositionTutorial} or contract-based implementation~\cite{delahaye2014robust,contractAssembly}, the safeguard mechanism usually remains outcome-enforcing: a regulator or central planner imposes feasibility, allocation, or performance requirements on the final decision. 
This motivates regulated coordination mechanisms that pursue system-level safeguards by shaping the negotiation process itself, rather than by directly solving or enforcing the final system-wide decision.}

\paragraph{Hybrid and Oversight Mechanisms}
Hybrid oversight mechanisms seek to combine decentralized autonomy with desirable global behavior through safety interventions, reward shaping, taxation-like incentives, or robustness-enhancing control signals~\cite{orseau2016interruptible,zhijian2023nash,akbar2022robust}. 
\fix{Literature in energy and infrastructure networks similarly uses pricing or reward signals to align local decisions with system-level constraints and service objectives~\cite{zhou2017incentive,harsh2025stochastic}.}
\fix{In these schemes, an external signal modifies local rewards, penalties, prices, or constraints so that agents remain autonomous while their effective objectives are biased toward safer, more robust, or more efficient behavior. 
This philosophy is closely aligned with our framework, where oversight does not replace agent-level decision making but regulates the conditions under which decentralized decisions are made.}

\fix{However, existing guarantees are typically tied to specific formulations, such as safe interruption~\cite{orseau2016interruptible}, equilibrium selection~\cite{zhijian2023nash}, robustness of learning dynamics~\cite{akbar2022robust}, or infrastructure service constraints~\cite{zhou2017incentive,harsh2025stochastic}. 
They do not directly address regulated noncooperative consensus, where agents must agree on a common operational candidate while the regulator controls intervention through implicit coordination signals. 
In particular, these works do not characterize the practical autonomy--regulation tradeoff by providing explicit bounds that link intervention strength to finite-round termination, convergence speed, and system-level performance.}

\paragraph{\fix{Multi-Agent Coordination in Air Traffic Management}}
\fix{Air traffic management (ATM) has long addressed multi-agent coordination under shared capacity, safety, and fairness constraints. 
Traditional approaches rely on centralized optimization, traffic-flow regulation, or rule-based allocation to manage aggregate demand and preserve operational feasibility~\cite{centralized_1,centralized_2,centralized_TMA,market_castelli}. 
As ATM decision making became more distributed across stakeholders, including air navigation service providers and airlines, collaborative decision making (CDM) emerged to incorporate operator preferences into flow-management decisions. 
Within CDM, mechanisms such as the Collaborative Trajectory Options Program, rationing-based allocation, and trajectory-option planning coordinate flight operators through shared constraints, priority rules, and structured preference submission~\cite{li2017ctop,cdm}. 
These approaches provide important operational foundations, but they typically retain centralized allocation logic or assume structured collaboration among stakeholders.}

\fix{Several methodological frameworks extend collaborative ATM coordination. 
Market-based mechanisms use auctions, prices, or payment rules to allocate scarce air traffic resources while accounting for operator preferences~\cite{market_castelli,market_brugnara,market_mehta,market_privacy}. 
Negotiation- and auction-based methods coordinate agents through bids, trades, or locally generated options and have been used in multi-agent task allocation and resource-management problems~\cite{cbba,amirkhani2022consensus,taco,bertsekas1992auction}. 
Game-theoretic approaches model strategic interactions among self-interested airspace users~\cite{im2024coordination,2023decentralized,im2025game}. 
Together, these methods move beyond purely centralized assignment by incorporating agent-level preferences, but they primarily address allocation, payment design, or decentralized coordination without explicitly regulating system-level outcomes.}

\fix{Coordination problems arise across strategic and tactical ATM operations, including flight-plan coordination~\cite{im2025game}, ground-delay programs~\cite{gdp}, terminal-area management~\cite{centralized_TMA}, approach sequencing~\cite{im2024coordination}, tactical conflict resolution~\cite{2023decentralized,tactical}, and emerging AAM routing and scheduling~\cite{surya_aam}. 
Although these problems differ in time scale and operational detail, they increasingly share a common feature: decision authority and operational information are distributed across multiple stakeholders. 
This trend motivates coordination mechanisms that preserve operator autonomy and private preferences while maintaining system-level safety, fairness, and efficiency.}

\section{Preliminaries: Decentralized noncooperative negotiation via \texttt{TACo}}
\label{sec:noncoop}

We employ the \emph{trading auction for consensus} (\texttt{TACo}) as a baseline mechanism for noncooperative negotiation \cite{taco}.
\texttt{TACo} enables self-interested stakeholders to agree on a common decision while pursuing different objectives and withholding private information.

\texttt{TACo} operates as a structured \emph{give-and-take} process, where agents iteratively trade a secondary asset (\eg, emission credits or monetary tokens) to persuade others to accept their preferred choice. It provides a fully decentralized procedure through which consensus naturally emerges from agents’ self-interested behavior and yields an outcome that remains incentive-aligned for all participants, without centralized coordination or direct bilateral communication.

Unlike standard single-round voting-based consensus mechanisms, \texttt{TACo} introduces a dynamic trading structure that aligns incentives through self-interested behavior. Although there is no formal guarantee, this mechanism often yields outcomes with a lower overall system cost and improved fairness compared to voting, as trading redistributes the burden among agents with differing valuations.

\subsection{Mechanism overview}

\texttt{TACo} is a fully decentralized mechanism in which agents follow the common trading rules and use only broadcast communication. Each agent $i \in \plSet := \{1, \ldots, n\}$ evaluates a finite set of options $\fix{\ell} \in \chSet := \{1, \ldots, m\}$, and computes its \textit{profit} $\fix{J_{i\ell}}$ for each option $\fix{\ell}$ as
\begin{equation} \label{eq:tacoProfitCalc}
    \fix{J_{i\ell}} = b_i (\fix{O_{i\ell}} - \fix{P_{i\ell}}) - \fix{C_{i\ell}},
\end{equation}
where:
\begin{itemize}
    \item $b_i \in \realSet_{\geq 0}$ is agent~$i$’s private \textit{valuation} of the traded secondary asset, which remains undisclosed to others.
    \item $\fix{C_{i\ell}} \in \realSet$ is the \textit{intrinsic cost} incurred by agent~$i$ when selecting option~$\fix{\ell}$.
    \item $\fix{O_{i\ell}} \in \realSet_{\geq 0}$ and $\fix{P_{i\ell}} \in \realSet_{\geq 0}$ denote the \textit{offer} and \textit{payment units}, representing the asset units received and paid, respectively, when agent $i$ selects option $\fix{\ell}$.
\end{itemize}
At each step, agent~$i$ takes its turn by selecting a profit-maximizing option, $\fix{\ell_i^*} \in \arg\max_{\fix{\ell} \in \chSet} \fix{J_{i\ell}}$, and broadcasting the selected option to all others.  
Upon this selection, the common offer and pay units are updated according to
\begin{align} \label{eq:tacoUpdate}
    \fix{P_{i\ell_i^*}} &\leftarrow \fix{P_{i\ell_i^*}} + n d, \quad \forall i \in \plSet,\\
    \fix{O_{i\ell_i^*}} &\leftarrow \fix{O_{i\ell_i^*}} + d, \quad \forall i \in \plSet,
\end{align}
where the current trading unit $d \in \realSet_{> 0}$ specifies the granularity of trading and $n$ is the number of agents.  
Given the broadcast information, these updates can be computed independently by all agents.

A \emph{cycle} occurs when the profit $J$ is observed to be the same by any agent $i$ at two different rounds of the negotiation---in this case, agent $i$'s (and all subsequent agents') choice of profit-maximizing options repeats.
Such a cycle indicates failure to reach a consensus with the current granularity of trading, $d$. Therefore when a cycle is detected, all agents reduce the trading unit via $d \leftarrow \gamma d$, where $\gamma \in (0,1)$, which leads to finer adjustments and smaller profit differences across options. The procedure repeats until all agents choose the same choice or the profit gap among all options in a cycle falls below a tolerance $\varepsilon \in \realSet_{>0}$, at which point agents become effectively indifferent among the choices and a consensus is declared. This procedure is guaranteed to terminate within a finite number of steps \cite{taco}.

\subsection{Properties of \texttt{TACo}}

The properties summarized below make \texttt{TACo} a practical foundation for decentralized coordination:

\paragraph{Incentive-aligned process}
Each agent acts \fix{according to its own asset valuation and induced preferences within the \texttt{TACo} negotiation}, and consensus emerges as the natural consequence of individual trading decisions rather than cooperative agreement. The resulting outcome also remains \fix{preference-sensitive within the negotiation stage}.

\paragraph{\fix{Non-disclosure of asset valuation}}
\texttt{TACo} does not require agents to directly disclose their \fix{private asset valuations $(b_i)$ during the negotiation process. 
In the proposed oversight framework, this non-disclosure property is preserved for asset valuations, while the information structure associated with operational cost quantities is specified in \Cref{sec:information_structure}}.

\paragraph{Decentralized mechanism}
\texttt{TACo} requires neither an auctioneer nor a central coordinator. All agents follow identical public rules and rely solely on broadcast information, enabling a fully decentralized implementation.

\paragraph{Finite termination guarantee}
The algorithm provably converges within a finite number of steps.

\vspace{0.3em} \noindent
For these reasons, we adopt \texttt{TACo} as the foundation of our negotiation framework. 

\subsection{Limitations of noncooperative negotiation without oversight}
\label{subsec:single-round-limit}

While \texttt{TACo} enables self-interested agents to reach consensus in a fully decentralized manner, it cannot change the underlying choices available to the agents, some or all of which may be poor. 
Furthermore, \texttt{TACo} does not optimize for any concept of system optimality, and is only guaranteed to find a solution that is unilaterally acceptable to all agents.

In safety-critical domains, these limitations motivate the need for an additional oversight mechanism that guides the negotiation toward system-level objectives. The following section introduces our main contribution: an iterative framework that embeds such an oversight into the \texttt{TACo}-based negotiation.

\section{Iterative negotiation and oversight framework}
\label{sec:framework}

\begin{figure}[hbt!]
    \centering
    \includegraphics[width=0.99\linewidth]{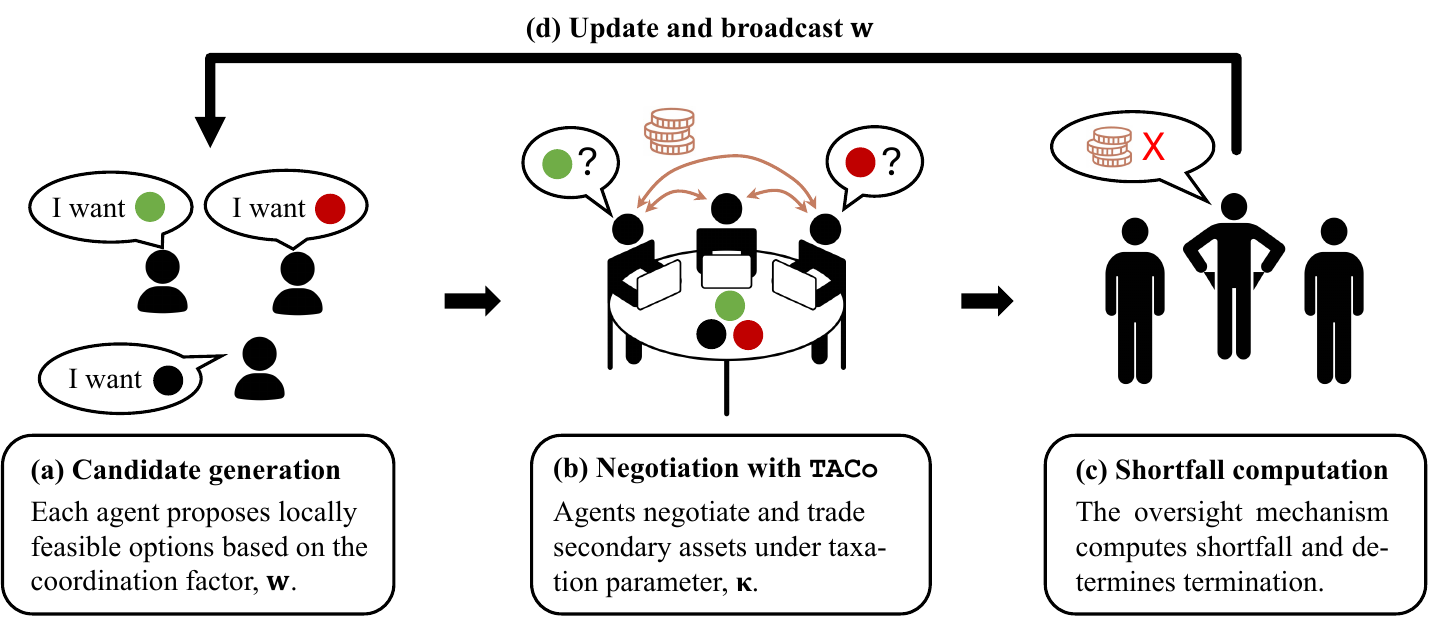}
    \caption{Overview of the iterative negotiation and oversight framework.
Each iteration consists of \fix{four} stages:
(a) Candidate generation: each agent proposes candidate options based on the broadcast coordination factor $\coordFactor$, (b) Negotiation with \texttt{TACo}: agents negotiate and trade secondary assets under the taxation parameter $\kappa$, and (c) Shortfall computation: the oversight mechanism computes shortfalls based on asset shortages and (d) broadcasts the updated coordination factor for the next round.}
    \label{fig:algOverview}
\end{figure}

We extend the \texttt{TACo} mechanism with an outer-loop---implemented by a central authority---\fix{that introduces regulatory guidance while preserving TACo's noncooperative negotiation stage}. Moreover, at each outer-loop round, we allow agents to propose new choices for negotiation, which are influenced by a coordination factor selected by the proposed mechanism based upon the previous rounds of negotiation. \fix{This structure is intended to model regulated decentralized coordination: agents retain the ability to generate and negotiate over candidate choices, while the oversight layer modifies the candidate-generation environment when previous negotiation rounds reveal system-level shortfalls.} 

Each outer iteration consists of four stages, as shown in \Cref{alg:oversight}: (i) candidate choice generation by each agent, (ii) a \texttt{TACo} negotiation based on the generated choices and the taxation-adjusted secondary asset values, (iii) shortfall evaluation with respect to trading expenditures and available asset reserves, and (iv) coordination factor update and broadcast. This four-step process (candidate–negotiation–shortfall–update) repeats until the termination condition is met.

\subsection{Framework flow}
\label{subsec:design}

\begin{algorithm}[hbt!]\label{alg:oversight}
\caption{Iterative negotiation and oversight framework}
\KwIn{Agent set $\plSet$, local action sets $\{\mathcal{A}_i\}_{i\in\plSet}$, asset reserves $\{R_i\}_{i\in\plSet}$, tax parameter $\kappa > 0$, initial coordination factor $\coordFactor^{(0)}$}

Initialize $r \gets 0$, $\coordFactor^{(0)}=\bm{0}^n$\;

\While{\textnormal{true}}{
    \nl \tcc{Candidate generation} \nllabel{alg:ln:candidate}
    \For{$i \in \plSet$}{
        Solve \Cref{eq:combinedCostNormalized} $\to$
        Generate a candidate $o_i^{(r)}$\;
    }
    Combine all local candidates into the joint candidate pool:
    $\opSet^{(r)} = \bigcup_{i \in \plSet} o_i^{(r)}$\;

    \vspace{0.3em}
    \nl \tcc{Noncooperative negotiation} \nllabel{alg:ln:nego}
    Execute \texttt{TACo}($\opSet^{(r)}$) $\rightarrow$ agreement choice $o^{(r)}$, payments $\{P_i\}_{i\in\plSet}$, and offers $\{O_i\}_{i\in\plSet}$\;

    \vspace{0.3em}
    \nl \tcc{Shortfall computation and termination check} \nllabel{alg:ln:shortfall}
    \For{$i \in \plSet$}{
        Compute expenditure $e_i = P_i - O_i$ and shortfall 
        $s_i = \frac{1}{R_i}\max\{0, e_i - R_i\}$\;
    }
    \If{$\sum_{i \in \plSet} s_i = 0$}{
        \textbf{break}\;
    }

    \vspace{0.3em}
    \nl \tcc{Coordination factor update} \nllabel{alg:ln:update}
    Normalize $\bar{s}_i = s_i / \sum_{j \in \plSet} s_j$ for all $i$\;
    Update coordination factor:
    $\coordFactor^{(r+1)} = \coordFactor^{(r)} + \bar{s}$\;
    $r \gets r + 1$\;
}
\Return{$o^* = o^{(r)}$}
\end{algorithm}

At the beginning of each round $r$, a central authority broadcasts a vector coordination signal $\coordFactor^{(r)} \in \realSet^n$. Each agent proposes a new candidate which satisfies its local constraints and optimizes an effective cost function that blends its local objective with the shared term scaled by the coordination factor, $\coordFactor^{(r)}$. Via \texttt{TACo}, the agents reach an agreement $o$ with associated payments $\{P_{io}\}_{i\in\plSet}$ and offers $\{O_{io}\}_{i\in\plSet}$, representing the amount of secondary assets traded per agent. The oversight mechanism then computes the \emph{shortfall} by comparing trade expenditures with each agent’s asset reserve and updates the coordination factor accordingly for the next round.

\subsection{\fix{Information structure and compliance assumptions}} 
\label{sec:information_structure}

\fix{The proposed framework is designed for a regulated coordination environment in which agents are self-interested but protocol-compliant. 
The central authority specifies the regulatory signal used in candidate generation, while candidate proposal and final selection remain decentralized through agent-side optimization and \texttt{TACo} negotiation.}

\fix{The candidate-generation cost is constructed from operational quantities that are shared among participants or computable by the regulator from information available within the coordination environment. 
This assumption is natural in regulated systems, where some system-state or feasibility-related quantities are observable to the coordinator independently of agents' private valuations or preferences.}

\fix{This yields a mixed information structure: operational quantities enter the regulated candidate-generation step, whereas the private asset valuations used in \texttt{TACo} remain undisclosed. 
Accordingly, the framework preserves non-disclosure of private asset valuations while using regulatory oversight to correct system-level deviations.}

\subsection{Candidate Choice Generation (\Cref{alg:oversight} \cref{alg:ln:candidate})}
\label{subsec:candidate}

\fix{This step represents the regulatory component of the framework. 
Each agent still generates its own candidate, but the effective cost used for candidate generation is prescribed by the oversight protocol and depends on the broadcast coordination factor.}

Recall that $\plSet = \{1, \dots, n\}$ denotes the set of agents, 
and that $\opSet$ denotes the set of possible individual candidate choices. 
At round~$r$, each agent~$i \in \plSet$ proposes a candidate choice $o_i^{(r)} \in \opSet$, 
and the set of candidates from all agents is represented by 
$\opSet^{(r)} = \{o_1^{(r)}, \dots, o_n^{(r)}\}$.

Define the \emph{individual cost function} $J^{\mathrm{ind}} : \opSet \to \mathbb{R}^n,$
\begin{equation}
    J^{\mathrm{ind}}(o)
    := \big[J_1^{\mathrm{ind}}(o), \dots, J_n^{\mathrm{ind}}(o)\big]^\top,
    \quad o \in \opSet,
\end{equation}
where $J_i^{\mathrm{ind}}(o) := [J^{\mathrm{ind}}(o)]_i$ 
represents the cost incurred by agent~$i$ under choice~$o$.

Given a nonnegative coordination factor at round~$r$, $\coordFactor^{(r)} \in \mathbb{R}_{\ge 0}^n$, \fix{which represents the regulatory weight assigned to agents' operational costs based on previous shortfalls,} the \emph{shared score} function $S_{\coordFactor^{(r)}} : \opSet \to \mathbb{R}$ is defined as
\begin{equation}\label{eq:sharedScore}
    S_{\coordFactor^{(r)}}(o)
    := (\coordFactor^{(r)})^\top J^{\mathrm{ind}}(o),
    \quad o \in \opSet.
\end{equation}

To control the relative weight between the local and shared terms, we normalize this cost function and use the normalized form as the basis for choice generation in every round. 
This normalization is instrumental in allowing us to prove that \Cref{alg:oversight} converges in a finite number of rounds.
Define the normalization factor $\alpha_r$ and its associated normalized coordination vector $\bar \coordFactor^{(r)}$ as
\begin{equation} \label{eq:normalizationFactor}
\alpha_r := \|\coordFactor^{(r)}\|_1 + 1,\qquad 
\bar{\coordFactor}^{(r)} := \frac{\coordFactor^{(r)}}{\alpha_r}.
\end{equation}
The normalized effective cost is then defined as
\begin{equation}\label{eq:combinedCostNormalized}
J^{\bar{\coordFactor}^{(r)}}_i(o)
\;:=\;
S_{\bar{\coordFactor}^{(r)}}(o)
+ \frac{1}{\alpha_r}\,J^{\mathrm{ind}}_i(o),
\qquad
S_{\bar{\coordFactor}^{(r)}}(o) = \frac{S_{\coordFactor^{(r)}}(o)}{\alpha_r}.
\end{equation}

\noindent
As $\alpha_r$ increases, the normalized shared term $S_{\bar{\coordFactor}^{(r)}}(o)$ gradually dominates, 
steering agents’ choices --- $o_i^{(r)}$ minimizing \Cref{eq:combinedCostNormalized} --- toward more coordinated and equitable outcomes, 
which, as shown later in \Cref{sec:conv-oversight}, implicitly leads to improved system-level efficiency. 
\fix{This increasing influence of the shared term captures the autonomy--oversight tradeoff: as additional rounds are required, the candidates available for negotiation become increasingly shaped by regulatory guidance, although the final selection among those candidates is still performed through \texttt{TACo}.}

\subsection{Negotiation with \texttt{TACo} (\Cref{alg:oversight} \cref{alg:ln:nego})}
\label{subsec:negotiation}

\fix{This stage constitutes the noncooperative negotiation component of the framework.}
Given agents' set of candidates $\opSet^{(r)}$, \texttt{TACo} yields an agreement and returns the amount of asset transfers that aligns with individual interests while \fix{avoiding direct disclosure of asset valuations.}
Let $R_i \in \realSet_{\geq 0}$ be agent $i$’s \textit{asset reserve}, \fix{representing the amount of secondary asset available for negotiation,} and $\kappa \in \realSet_{\geq 0}$ a \textit{tax parameter}.  
We employ an asset valuation method based on the \textit{constant-product market-making} principle~\cite{CPMM}:
\begin{equation} \label{eq:assetValuation}
b_i \;=\; \frac{1}{\kappa R_i}.
\end{equation}
The \texttt{TACo} then returns an agreement $o^{(r)}$ with payment $\{P_{io^{(r)}}\}_{i\in\plSet}$ and offer $\{O_{io^{(r)}}\}_{i\in\plSet}$ for each agent.

\begin{remark}[\fix{Interpretation of asset reserves}]
\fix{The reserve $R_i$ denotes the amount of secondary asset available to agent $i$ within the negotiation process. Depending on the implementation, this asset may represent monetary tokens, emission credits, or regulatory credits. Different reserve amounts are used to evaluate sensitivity to asset expenditures in the numerical experiments in Section~7, rather than to represent an agent's external market power, airline size, etc.}
\end{remark} \label{rem:R_interpretation}

\subsection{Shortfall computation and coordination factor update (\Cref{alg:oversight} \cref{alg:ln:shortfall} and \cref{alg:ln:update})}
\label{subsec:shortfall}

Given an agreement outcome $o^{(r)}$, we compute the \emph{expenditure} $e_i$ and the \emph{shortfall} $s_i$:
\begin{equation} \label{eq:shortfall}
e_i := P_{io^{(r)}} - O_{io^{(r)}},\quad
s_i := \frac{1}{R_i}\max\{0,\,e_i - R_i\}, \quad \forall i\in\plSet.
\end{equation}
The shortfall quantifies the amount of excess trading assets required for an agreement to be reached \fix{relative to the protocol-specific credit budget assigned to each agent}.  
When $\sum_{j\in\plSet} s_j=0$, the \emph{asset reserve constraint} is satisfied because all agents satisfy $e_i\leq R_i$; in this case, \Cref{alg:oversight} terminates.  
If any shortfall is detected ($\exists i \in \plSet$ such that $s_i>0$), the agents cannot settle in the current negotiation round, and an additional round is initiated.  
Define the normalized shortfall $\bar{s}_i := s_i / \sum_{j \in \plSet} s_j$ and update the coordination factor by
\begin{equation}
\coordFactor^{(r+1)} \;=\; \coordFactor^{(r)} + \bar{s}.
\end{equation}
This update increases the coordination weight for agents that exceed their reserves while leaving others unchanged. Under this rule, $\alpha_{r+1} = \alpha_r + 1$ because $\|\bar{s}\|_1 = 1$, 
indicating that the influence of the shared term increases monotonically across rounds, cf. \eqref{eq:combinedCostNormalized}. \fix{Thus, the shortfall update serves as a regulatory correction that links reserve violations in the current negotiation round to stronger coordination guidance in the next round.}

\subsection{Oversight role and safeguard}
\label{subsec:oversight}

As the shared term in the effective cost~\Cref{eq:combinedCostNormalized} gains greater influence across rounds, agents generate progressively more coordinated and equitable candidate choices, enabling consensus to be reached with smaller asset transfers. Such gradual coordination is also expected to implicitly contribute to improved system-level efficiency. 
\fix{This mechanism captures an autonomy--oversight tradeoff: when additional rounds are required, regulatory guidance becomes stronger at the expense of fully autonomous candidate generation, while \texttt{TACo} still selects among the generated candidates.}

We claim that the framework always converges to an outcome satisfying the reserve constraints within a finite number of rounds. Moreover, the taxation parameter~$\kappa$ governs both the convergence speed and the resulting system efficiency: a smaller~$\kappa$ leads to faster negotiations (fewer rounds) and \fix{weaker regulatory correction}, whereas a larger~$\kappa$ \fix{allows more rounds of oversight-guided candidate generation, which can promote more system-efficient outcomes} at the expense of slower convergence and reduced candidate-generation autonomy. These properties are formalized and proven in the following section.

\section{Convergence Proof and Oversight Guide}
\label{sec:conv-oversight}

We establish the convergence and performance guarantees of the proposed iterative negotiation with oversight. We first prove \emph{finite termination} of the framework, and then relate the tax parameter~$\kappa$ to (i) an upper bound for the number of rounds until convergence and (ii) a \emph{system-optimality gap} bound. 

\subsection{Finite-round convergence}
\label{subsec:conv-proof}

We now establish that the proposed iterative negotiation with oversight framework always converges within a finite number of rounds. Before presenting the theorem that guarantees convergence, we outline the key steps underlying the proof. First, we show that as $\|\coordFactor^{(r)}\|_1$, \ie, the magnitude of the coordination factor, increases, all agents eventually generate candidates with nearly equivalent costs (Step~1). Second, we show that this flattening bounds the magnitude of individual asset transfers between agents after \texttt{TACo} (Step~2). Finally, we prove that the process must terminate as all players eventually satisfy any nonzero asset reserves (Step~3).

\paragraph{Step 1: Choice-set flattening}
As the norm of coordination factor $\coordFactor^{(r)}$ grows, the normalized shared term dominates in the cost function~\eqref{eq:combinedCostNormalized}, and every agent’s perceived cost difference among candidate choices diminishes.

\begin{lemma}[Iterative choice generation yields convergent choice sets]
\label{lemma:choice-conv}
Recall that $\opSet$ denotes the complete set of feasible choices available to all agents\footnote{We assume that $\opSet$ is a finite set so that all extrema and summations in this section are well-defined.}, and $\opSet^{(r)}\subseteq\{1,\dots,m\}$ denotes the set of choices generated by all agents at round~$r$. At outer round $r=1,2,\dots$, recall that $\alpha_r := \|\coordFactor^{(r)}\|_1 + 1$ and that $\alpha_{r+1}=\alpha_r+1$.

For each agent~$i$, define the maximum spread of its intrinsic costs as
\begin{equation} \label{eq:bmax}
B_i := \max_{o,o' \in \opSet} \big|J_i^{\mathrm{ind}}(o) - J_i^{\mathrm{ind}}(o')\big|,
\qquad
B_{\max} := \max_i B_i.
\end{equation}
Then, for all $i$ and $r$,
\begin{equation}
\max_{o,o'\in\mathcal O^{(r)}} 
\big|J_i^{\coordFactor^{(r)}}(o)-J_i^{\coordFactor^{(r)}}(o')\big|
\;\le\; \frac{B_i}{\alpha_r}
\;\le\; \frac{B_{\max}}{\alpha_r}
\;\xrightarrow[r\to\infty]{} 0.
\end{equation}
As $\alpha_r$ increases, the shared term flattens each agent’s effective cost landscape,  
causing the within-agent cost spread to shrink at a rate proportional to $1/\alpha_r$.  
\begin{proof}
    The proof is provided in \ref{app1}.
\end{proof}
\end{lemma}

\paragraph{Step 2: Transfer value bounds}
We leverage \Cref{lemma:choice-conv} to show that the flattening of effective cost functions constrains the range of \emph{transfer values}—the \emph{value} of assets exchanged between agents—during each \texttt{TACo} negotiation.
As the effective costs flatten, the potential gains from trading diminish, and each agent required to expend its assets (payer) faces a transfer value that is bounded by a factor inversely proportional to $\alpha_r$.

\begin{lemma}[Individual transfer value bounds]
\label{lemma:ind-transfer}
Let $t_i^{(r)}$ denote agent $i$’s transfer value at round~$r$, 
and let $o^{\texttt{TACo}}$ be the outcome selected by \texttt{TACo}. 
Define
\begin{equation}
o_{i}^{(1)} := \arg\min_{o \in \opSet} J_i^{\coordFactor^{(r)}}(o),
\end{equation}
and let $L_r := \{ i : o^{\texttt{TACo}} \not\subset o_{i}^{(1)} \}$ the set of beneficiaries and 
$W_r := \{ i : o^{\texttt{TACo}} \subset o_{i}^{(1)} \}$ the set of payers. 
Then there exist nonnegative $u_i,\fix{v_j}$ such that
\begin{subequations}
\begin{align}
    0 \leq u_i \le |t_i^{(r)}|, &\quad \forall i\in L_r, \label{eq:transfer-lower}\\
    0 \leq |t_j^{(r)}| \le \fix{v_j}, &\quad \forall j\in W_r. \label{eq:transfer-upper}
\end{align}
\end{subequations}
\begin{proof}
    The proof is provided in \ref{app1}.
\end{proof}
\end{lemma}

The inequalities \Cref{eq:transfer-lower} and \Cref{eq:transfer-upper} formalize the transfer rules in \texttt{TACo}. A payer will not transfer an amount that would make its current choice less favorable than the next-best alternative, thereby ensuring the upper bound in~\Cref{eq:transfer-upper}. Conversely, a beneficiary must receive at least enough value to prefer the resulting outcome to its previously optimal choice, guaranteeing the lower bound in~\Cref{eq:transfer-lower}. As shown in \Cref{lemma:choice-conv}, the flattening of agents’ effective cost landscapes across rounds reduces the difference between each agent’s best and second-best options, which in turn tightens both the upper and lower bounds on transfer values. The next result establishes a quantitative upper bound on these values as a function of~$\alpha_r$.

\begin{corollary}[Payment bound for payers]
\label{cor:round-bound}
The transfer value for every payer satisfies
\begin{equation}
0 \le |t_j^{(r)}| \le \frac{B_{\max}}{\alpha_r},\quad \forall j \in W_r,
\end{equation}
where $B_{\max}$ is the maximum spread of intrinsic choice costs as defined in \Cref{lemma:choice-conv}.
\begin{proof}
    The proof is provided in \ref{app1}.
\end{proof}
\end{corollary}
As the shared term in \Cref{eq:combinedCostNormalized} dominates, the upper bound on each transfer value decreases with~$1/\alpha_r$,  
implying that the total traded value becomes progressively smaller as the process evolves.

\paragraph{Step 3: Finite termination}
We now incorporate the quantization of asset transfers and the asset reserves $R_i\in\realSet_{>0}$ to show that the process must terminate after a finite number of rounds.  
Because each transfer is quantized into discrete units and each agent’s reserve is finite,  
the process halts once all payers reach their unit-count limits.

\begin{theorem}[Finite termination under unit-count reserves]
\label{theorem:termination-ind}
Assume that the asset transfer values are quantized as $|t_j^{(r)}| = c_j^{(r)} b_j$, 
where $b_j = \frac{1}{\kappa R_j}$ from \Cref{eq:assetValuation} and $c_j^{(r)} \in \intSet_{\ge 0}$ denotes the trading quantity in fixed asset units.  
Then there exists a finite round $\bar{r}$ such that $c_j^{(r)} \le R_j$ for all $r \ge \bar{r}$ and $j \in W_r$.  
Hence, the process terminates once all agents satisfy their asset-reserve constraints, 
i.e., no agent pays more than the amount available in its reserve.
\begin{proof}
    The proof is provided in \ref{app1}.
\end{proof}
\end{theorem}

\noindent
When a negotiation round fails due to excessive asset expenditure,  
the framework initiates another round, increasing $\alpha_r$ and further flattening the effective costs. As established in the previous steps, this reduces an upper bound on the amount of asset transfers required to reach consensus through \texttt{TACo}. Consequently, each payer’s expenditure eventually falls below its reserve $R_i$, guaranteeing that the process terminates within a finite number of rounds.

\subsection{The relationship between the taxation parameter $\kappa$ and convergence rate}
\label{subsec:kappa-rate}

We now relate the convergence rate---the number of rounds required until termination---to the taxation parameter~$\kappa$. Intuitively, $\kappa$ regulates the value of the trading units $b_i = 1/(\kappa R_i)$ and thereby controls the amount of assets exchanged during negotiation, influencing how quickly the reserve constraints are satisfied. For example, a smaller $\kappa$ (larger asset values) induces larger trading steps, which means that agents can reach consensus while trading fewer assets (not depleting their reserves), and ultimately converge more quickly.

\begin{theorem}[Sufficient condition for termination in terms of $\kappa$] 
\label{theorem:tax-to-alpha}
If $\alpha_r \ge \kappa B_{\max}$ at some round $r$, then every payer respects its asset reserve constraint ($c_j^{(r)} \le R_j$ for all $j\in W_r$), and the process terminates.
\begin{proof}
    The proof is provided in \ref{app1}.
\end{proof}
\end{theorem}

\noindent
\Cref{theorem:tax-to-alpha} indicates that termination is guaranteed once the normalization factor $\alpha_r$ exceeds a threshold determined by the product of the tax parameter~$\kappa$ and the upper bound on individual transfer values~$B_{\max}$ defined in \Cref{lemma:choice-conv}.  
This provides a direct link between $\kappa$ and $r_{\mathrm{term}}$, the number of rounds required for convergence.

\begin{corollary}[Upper bound on termination round]
\label{corollary:round-upper}
With $\alpha_1=1$ and $\alpha_r=r$, the process must terminate once $r \ge \kappa B_{\max}$, i.e.,
\begin{equation}
r_{\mathrm{term}} \;\le\; \big\lceil \kappa B_{\max} \big\rceil.
\label{eq:round-upper}
\end{equation}
Equivalently, for a desired iteration budget $r_{\mathrm{des}}\in\mathbb{N}$, the process is guaranteed to terminate if
\begin{equation}\label{eq:round-kappa}
\big\lceil \kappa B_{\max} \big\rceil \;\le\; r_{\mathrm{des}}
\quad\Longleftrightarrow\quad
\kappa \;\le\; \frac{r_{\mathrm{des}}}{B_{\max}}.
\end{equation}
\begin{proof}
    The proof is provided in \ref{app1}.
\end{proof}
\end{corollary}

\noindent
This result explicitly shows that $\kappa$ acts as a tuning parameter that regulates the convergence rate: smaller $\kappa$ values yield faster termination, while larger $\kappa$ values provide finer equilibrium adjustments that take place over more rounds \fix{and allow more opportunity for regulatory guidance to influence subsequent candidate-generation steps}.

\subsection{The relationship between the taxation parameter $\kappa$ and system efficiency}
\label{subsec:kappa-eff}

We now analyze how $\kappa$ affects the \emph{system-level efficiency} of the resulting consensus, 
that is, the gap between the achieved outcome and the global optimum. Intuitively, a larger $\kappa$ leads to a greater number of rounds and increases the effect of the coordination factor. This promotes improved fairness among agents, which in turn leads to outcomes closer to the system optimum, at the expense of more rounds. 

We first formalize the definition of \emph{system-optimal outcome}. 
\begin{definition}[System-optimal outcome]
    Let $u = \tfrac{1}{n}\mathbf{1}^n$ denote a uniform weight vector, where $\mathbf{1}^n = [1,\cdots,1] \in \realSet^n$. We define the \emph{system-level cost} as
    \begin{equation}
    J_{\mathrm{sys}}(o) := \sum_{i=1}^n J_i^{\mathrm{ind}}(o) = n S_u(o),
    \end{equation}
    where $S_u(o) = u^\top J^{\mathrm{ind}}(o)$ as defined in \Cref{eq:sharedScore}. A corresponding \emph{system-optimal choice} is
    \begin{equation}
    o^{\mathrm{opt}} \in \arg\min_{o \in \opSet} J_{\mathrm{sys}}(o).
    \end{equation}
\end{definition}
\noindent
Thus, the objective of the oversight is to steer the negotiation outcome $o_r^{\texttt{TACo}}$ in a  way that makes $J_{sys}(o_r^{\texttt{TACo}})$ closer to $J_{sys}(o^{\mathrm{opt}})$. 

\paragraph{Step 1: Selection error}
The following lemma bounds the \emph{selection error}, that is, the deviation between the selected choice and the one minimizing the normalized shared score $S_{\bar\coordFactor^{(r)}}(o)$.

\begin{lemma}[Selection error bound]
\label{lemma:eta-bound}
Fix round $r$ and let $o^\dagger \in \arg\min_o S_{\bar\coordFactor^{(r)}}(o)$. Then
\begin{equation}
0 \le 
\underbrace{S_{\bar\coordFactor^{(r)}}(o_r^{\texttt{TACo}})-S_{\bar\coordFactor^{(r)}}(o^\dagger)}_{\displaystyle \eta_r}
\le \frac{B_{\max}}{\alpha_r}.
\end{equation}
\begin{proof}
    The proof is provided in \ref{app1}.
\end{proof}
\end{lemma}

\noindent
As $\alpha_r$ increases, the selection error $\eta_r$ shrinks to zero.

\paragraph{Step 2: System optimality gap} 
Building on the selection error bound in \Cref{lemma:eta-bound}, we now bound the system–optimality gap as a function of the normalization factor~$\alpha_r$.
\begin{lemma}[Relationship between $\alpha$ and system–optimality gap bound]
\label{lem:sys-gap}
Recall that $J_{\mathrm{sys}}(o)=nS_u(o)$ with $u=\frac{1}{n}\mathbf{1}^n$ and $\bar \coordFactor^{(r)}=\coordFactor^{(r)}/\alpha_r$. Let $\delta_r:=\|\bar \coordFactor^{(r)}-u\|_1$, then
\begin{equation} \label{eq:gap-bound-by-round}
0 \le J_{\mathrm{sys}}\big(o_r^{\texttt{TACo}}\big)-J_{\mathrm{sys}}(o^{\mathrm{opt}})
\le n \left( B_{\max}\,\delta_r + \eta_r \right)
\le n \left( B_{\max}\,\delta_r + \frac{B_{\max}}{\alpha_r} \right).
\end{equation}
\begin{proof}
    The proof is provided in \ref{app1}. 
\end{proof}
\end{lemma}

\noindent
This lemma shows that the overall system inefficiency is governed by two quantities:  
the \emph{weight misalignment} $\delta_r$ between agents’ effective coordination weights and the uniform distribution,  
and the \emph{selection error} $\eta_r$ from \Cref{lemma:eta-bound}.  
As $\alpha_r$ increases, the selection error term diminishes, while the misalignment term decreases as $\bar\coordFactor^{(r)}$ approaches the uniform weight $u$.
Ultimately, as $\alpha_r$ diminishes the $\frac{B_{\max}}{\alpha_r}$, the system efficiency is bounded by $nB_{\max}\delta_r$.

\paragraph{Step 3: A necessary $\kappa$ for a target gap}
We finally translate the previous $\alpha_r$–optimality gap relationship into a necessary condition on~$\kappa$ to satisfy a target optimality gap, $\tau$. From \Cref{eq:normalizationFactor}, note that the $\alpha_r$ increases by one at each round and is initialized as $\alpha_1 = 1$, so $\alpha_r = r$. 
\begin{theorem}[Necessary $\kappa$ for a target optimality gap]
\label{thm:necessary-kappa}
Let $\alpha_r=r$ and $r_{\max}:=\lceil \kappa B_{\max}\rceil$, cf. \Cref{corollary:round-upper}. Assume $\delta_r\le \bar\delta$ for all $r\le r_{\max}$ and let $\tau>0$ denote the desired optimality gap. To certify at termination that
\begin{equation} \label{eq:optGap-upper}
J_{\mathrm{sys}}\big(o_{r_{\mathrm{term}}}^{\texttt{TACo}}\big)-J_{\mathrm{sys}}(o^{\mathrm{opt}})\le \tau,
\end{equation}
it is \emph{necessary} that $\tau>nB_{\max}\bar\delta$ and
\begin{equation}
r_{\max} \;\ge\; 
\frac{n B_{\max}}{\ \tau - n B_{\max}\bar\delta\ },
\quad\text{equivalently}\quad
\kappa \;\ge\; \frac{1}{B_{\max}}\left\lfloor \frac{nB_{\max}}{\ \tau-nB_{\max}\bar\delta\ }\right\rfloor.
\end{equation}
\begin{proof}
    The proof is provided in \ref{app1}.
\end{proof}
\end{theorem}

\begin{corollary}[Eliminating $\bar\delta$ to obtain a practical bound]
\label{cor:kappa-geometry}
Since $\delta_r=\|\bar\coordFactor^{(r)}-u\|_1 \le \sigma_{\max}$ with $\sigma_{\max}=2(1-\tfrac{1}{n})$, plugging $\bar\delta=\sigma_{\max}$ into the previous theorem (valid whenever $\tau>2(n-1)B_{\max}$) yields
\begin{equation} \label{eq:optGap-kappa}
\kappa \;\ge\; 
\frac{1}{B_{\max}}
\left\lfloor \frac{nB_{\max}}{\ \tau - 2(n-1)B_{\max}\ }\right\rfloor.
\end{equation}
\begin{proof}
    The proof is provided in \ref{app1}.
\end{proof}
\end{corollary}
\noindent
System efficiency improves with (i) better weight alignment ($\delta_r\downarrow$)---reflecting the balance in agents’ coordination influence---and (ii) more rounds ($\alpha_r\uparrow$).
In practice, misalignment often stems from uneven reserve distributions: agents with larger reserves tend to experience smaller shortfalls and thus contribute less influence in $\coordFactor$, whereas those with smaller reserves contribute more, amplifying the misalignment across rounds.

\subsection{Concise guide for the central coordinator}
We combine the analytical results derived in this section to provide a quick-reference guide 
for central coordinators to decide which $\kappa$ value to select in order to achieve a desired system behavior.

\begin{enumerate}
    \item To ensure that the negotiation process terminates within $r_{\mathrm{des}}$ rounds, 
    choose $\kappa \leq \frac{r_{\mathrm{des}}}{B_{\max}}$, cf. \Cref{eq:round-kappa}.
    \item To enable the possibility of achieving a target optimality gap $\tau$, 
    choose $\kappa \geq \frac{1}{B_{\max}} 
    \left\lfloor \frac{nB_{\max}}{\ \tau - 2(n-1)B_{\max}\ }\right\rfloor$, cf. \Cref{eq:optGap-kappa}.
    \item To guarantee a system optimality gap smaller than $\tau$, 
    enforce the number of rounds to satisfy 
    $r_{\mathrm{des}} \geq \frac{nB_{\max}}{\tau - nB_{\max}\delta_r}$, cf. \Cref{eq:gap-bound-by-round}.
\end{enumerate}

\paragraph{Practical remarks}
\begin{enumerate}
    \item The target optimality gap $\tau$ cannot be smaller than $nB_{\max}\delta_r$.
    \item There exists an intrinsic trade-off between convergence speed and system efficiency.
    Once $r_{\mathrm{des}}$ is fixed, it limits the range of achievable $\tau$, 
    and vice-versa.
\end{enumerate}

\subsection{\fix{Operational remarks}}
\fix{Beyond the choice of $\kappa$, the following remarks clarify how the analytical results should be interpreted in terms of scalability, autonomy, and privacy.}

\begin{remark}[\fix{Scalability by framework stage}]
\fix{Let $n$ denote the number of agents, $m_r := |\opSet^{(r)}|$ the number of generated candidates at outer round $r$, and $r_{\mathrm{term}}$ the number of outer-loop rounds. 
The shortfall computation and coordination-factor update scale as $O(n)$ per outer round, yielding $O(nr_{\mathrm{term}})$ total oversight-update overhead with $r_{\mathrm{term}}\leq \lceil \kappa B_{\max}\rceil$. 
According to the termination analysis of \texttt{TACo}~\cite{taco}, the inner negotiation step bound has a logarithmic dependence on the trading-unit parameters, approximately of the form $\left\lceil \log_\gamma(\varepsilon/d_0)\right\rceil$ up to problem-dependent scaling terms. 
With respect to \(n\) and \(m_r\), prior empirical studies have shown that \texttt{TACo} converges far more quickly than the theoretical bounds suggest \cite{taco}.
In the present framework, the measured runtime is dominated by candidate choice generation (cf. \Cref{subsec:candidate}), while the oversight update and \texttt{TACo} negotiation account for a comparatively small portion of the total computation time; the runtime decomposition by framework stage is reported in \Cref{subsec:runtime}.}
\end{remark} \label{sec:scalability}

\begin{remark}[\fix{Interpretation of the autonomy--oversight tradeoff}]
\fix{The number of outer-loop rounds has an autonomy and privacy interpretation. 
If the framework terminates in the first round, the final outcome is selected by the initial \texttt{TACo} negotiation without additional coordination-factor updates. 
In this case, the process remains closest to the baseline \texttt{TACo} setting: private asset valuations $b_i$ are not directly disclosed, and no additional information leakage arises from repeated oversight signals.
When additional rounds are required, the oversight layer accumulates shortfall information and modifies subsequent candidate generation. 
This can improve system-level outcomes, but it weakens the autonomy of candidate generation and may leak information about agents' private asset valuations \(b_i\) through the evolution of shortfalls \(s_i^{(r)}\) and coordination signals \(w_i^{(r)}\) across multiple rounds.
Therefore, the coordinator should select $\kappa$ with this autonomy--oversight tradeoff in mind: larger $\kappa$ may allow stronger system-level correction, but it also moves candidate generation further away from purely autonomous choices.}
\fix{A natural extension is to learn or adapt $\kappa$ across outer-loop rounds, using weaker intervention to encourage rapid initial agreement and stronger intervention when system-level improvement is insufficient, indicating the need for additional oversight. This direction is left for future work.}
\end{remark}

\section{Case study: Decentralized collaborative trajectory options program}
\label{sec:dctop}
\begin{figure}[t!]
    \centering
    \includegraphics[width=0.99\linewidth]{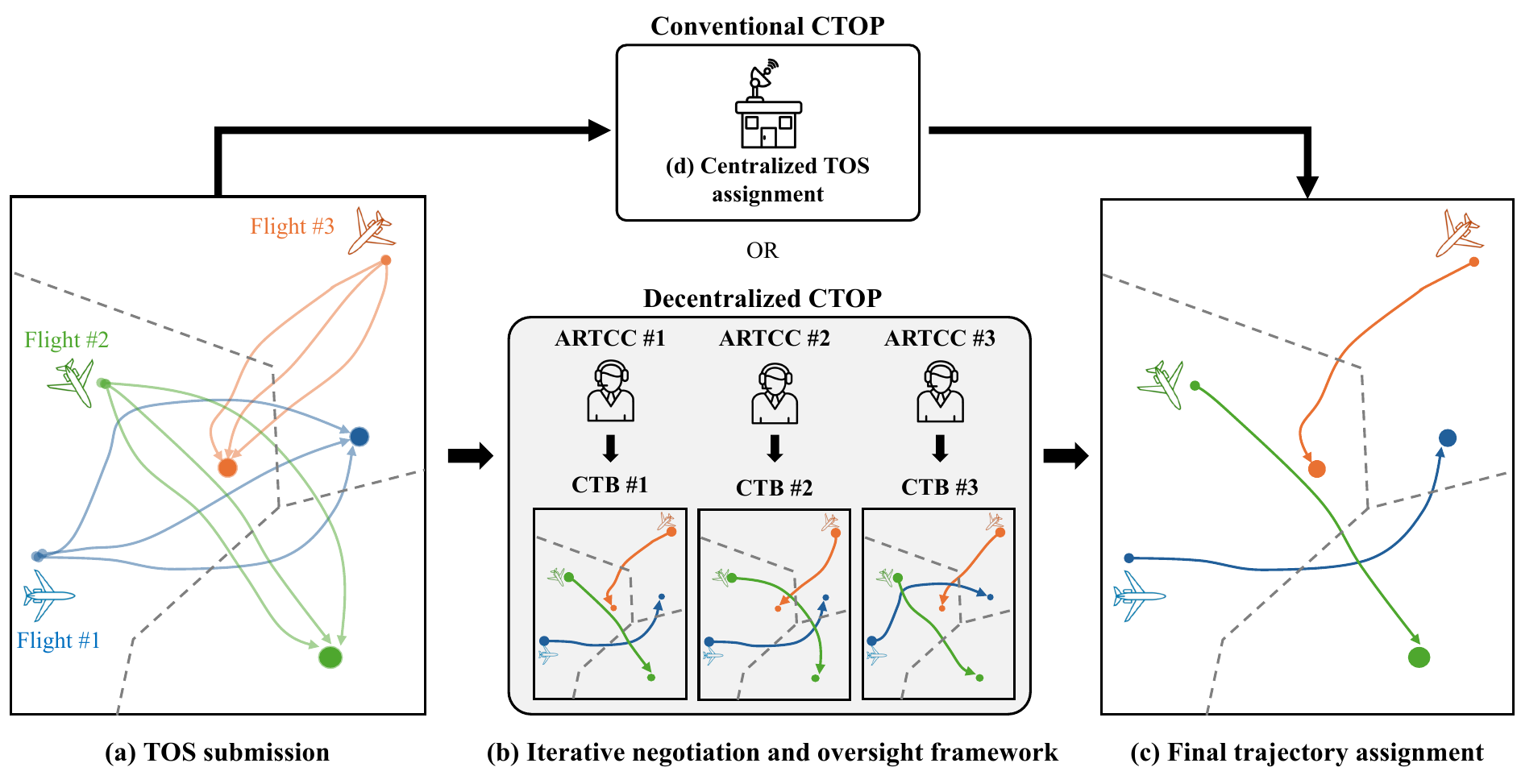}
    \caption{
    Comparison between the conventional and decentralized collaborative trajectory options program (CTOP).
    (a) Each airline submits a set of trajectory option sets (TOS) that differ in route, altitude, or arrival time.
    (b) In the proposed \fix{regulated decentralized} CTOP, each air route traffic control center (ARTCC) operates as an autonomous agent that generates candidate trajectory bundles (CTBs) and engages in iterative negotiation and oversight to coordinate with neighboring centers.
    (c) The final trajectories are determined through this decentralized mechanism, achieving coordinated outcomes \fix{without direct centralized trajectory assignment}.
    (d) In the conventional CTOP, a centralized air traffic control command center assigns trajectories based on predicted sector demand and capacity constraints.}
    \label{fig:dctop}
\end{figure}
Having established the theoretical framework for decentralized negotiation and oversight,
we now demonstrate its practical application to the \textit{collaborative trajectory options program} (CTOP).
This case study illustrates how the proposed framework can \fix{introduce regulated decentralization into} the existing air traffic management operation paradigm, which traditionally involves centralized decision-making. The overall flow of both the conventional and \fix{regulated decentralized} CTOP frameworks 
is summarized in \Cref{fig:dctop}. 

\subsection{Collaborative trajectory options program (CTOP)}
\label{subsec:ctop}

The CTOP is a traffic management initiative designed in the United States to mitigate airspace congestion by coordinating trajectory options for multiple flights \cite{faa_ac90115,canso_cdm,rodionova2017ctop}. In CTOP, each airline submits a set of alternative flight trajectories, referred to as \emph{trajectory options sets} (TOS), where each flight is associated with $q$ trajectory options that differ in route, altitude, or estimated arrival time. A centralized decision-making entity evaluates these submitted trajectories based on predicted sector demand, weather conditions, and sector capacity constraints. The central controller then assigns a trajectory to each flight, balancing global efficiency with equity among airlines.

\subsection{Decentralized CTOP}
\label{subsec:dctop-overview}

While CTOP enhances operational flexibility by considering the preferences of each individual airline, it does not account for the preferences of the individual airspace sector managers---or \emph{air route traffic control centers} (ARTCCs). Thus, the current paradigm still presents a potential for further decentralization, given the scalability demands of CTOP and the practical limits on cross-center information sharing \cite{chin2024centralized}. Building on the negotiation and oversight framework introduced in \Cref{sec:framework}, we develop a decentralized formulation of the CTOP, 
where each airspace sector, ARTCC, is modeled as a self-interested \emph{agent} responsible for assigning trajectory options for all flights traversing its region to minimize its own operational congestion. 
Unlike the centralized CTOP, where a single decision maker determines the global schedule, the decentralized CTOP allows sectors to suggest their preferred bundles of trajectory options and negotiate to reach a feasible agreement \fix{under regulatory oversight}.

\subsubsection{Candidate trajectory bundle generation}
\label{subsubsec:bundle}

Each ARTCC~$i \in \plSet$ corresponds to an agent in the proposed framework. 
All agents operate over the same set of $p$ flights, 
and each agent~$i$ constructs a \emph{candidate trajectory bundle} (CTB) 
that represents its preferred combination of trajectory options for the flights. 
A CTB for agent $i$ is expressed as a vector $o_i = [o_{i1}, \ldots, o_{ip}] \in \{1,\dots,q\}^p$, 
where each element $o_{if}$ denotes the trajectory option selected for flight~$f$.  
The common set of all possible CTBs is denoted by $\opSet \subseteq \{1,\dots,q\}^p$, and the collection of CTBs selected by all agents at round~$r$ is $\opSet^{(r)} = \{o_1^{(r)}, \ldots, o_n^{(r)}\}$.

Each agent~$i$ evaluates the congestion cost of a bundle $o_i \in \opSet$ using an individual cost function $J_i^{\mathrm{ind}} : \opSet \to \mathbb{R}$:
\begin{equation}
    J_i^{\mathrm{ind}}(o_i) = \texttt{EigenComplexity}(o_i).
\end{equation}
$\texttt{EigenComplexity}(\cdot)$ is a \fix{protocol-defined} scalar congestion metric computed 
from the leading eigenvalue of the sector traffic covariance matrix~\cite{eigenComplexity}, 
providing a quantitative measure of sector-level traffic complexity.
\fix{In this case study, this metric is treated as an operational quantity computable from shared traffic-flow information, rather than as a private cost report from each ARTCC.} 
Each agent constructs its CTB~$o_i^{(r)}$ under the effective cost formulation defined in \Cref{eq:combinedCostNormalized}.

\subsubsection{Noncooperative negotiation and asset valuation}
\label{subsubsec:nego}

Given the set of candidate bundles $\opSet^{(r)} = \{o_1^{(r)}, \ldots, o_n^{(r)}\}$ at round $r$, 
the $n$ ARTCC agents engage in a \texttt{TACo}-based negotiation to determine 
a mutually agreed combination of trajectory options. 
Each agent participates in \texttt{TACo} backed by its asset reserves~$R_i$, 
which represent a pool of \fix{regulatory tokens used within the negotiation process, such as emission credits, rather than a measure of an agent's external market power or airline size}.
Recall that the asset valuation is defined as
\begin{equation}
    b_i = \frac{1}{\kappa R_i},
\end{equation}
where $\kappa$ is the global taxation parameter broadcast by the oversight mechanism per \Cref{eq:assetValuation}. \fix{As in the general framework, the valuation $b_i$ is not directly disclosed during negotiation.} After each negotiation round, the oversight mechanism evaluates reserve shortfalls (cf. \Cref{eq:shortfall}) and updates the coordination factor $\coordFactor^{(r)}$ according to the shortfall-based rule described in \Cref{subsec:shortfall}. 

\section{Numerical experiment}

\begin{figure}[hbt!]
    \centering
    \includegraphics[width=0.6\linewidth]{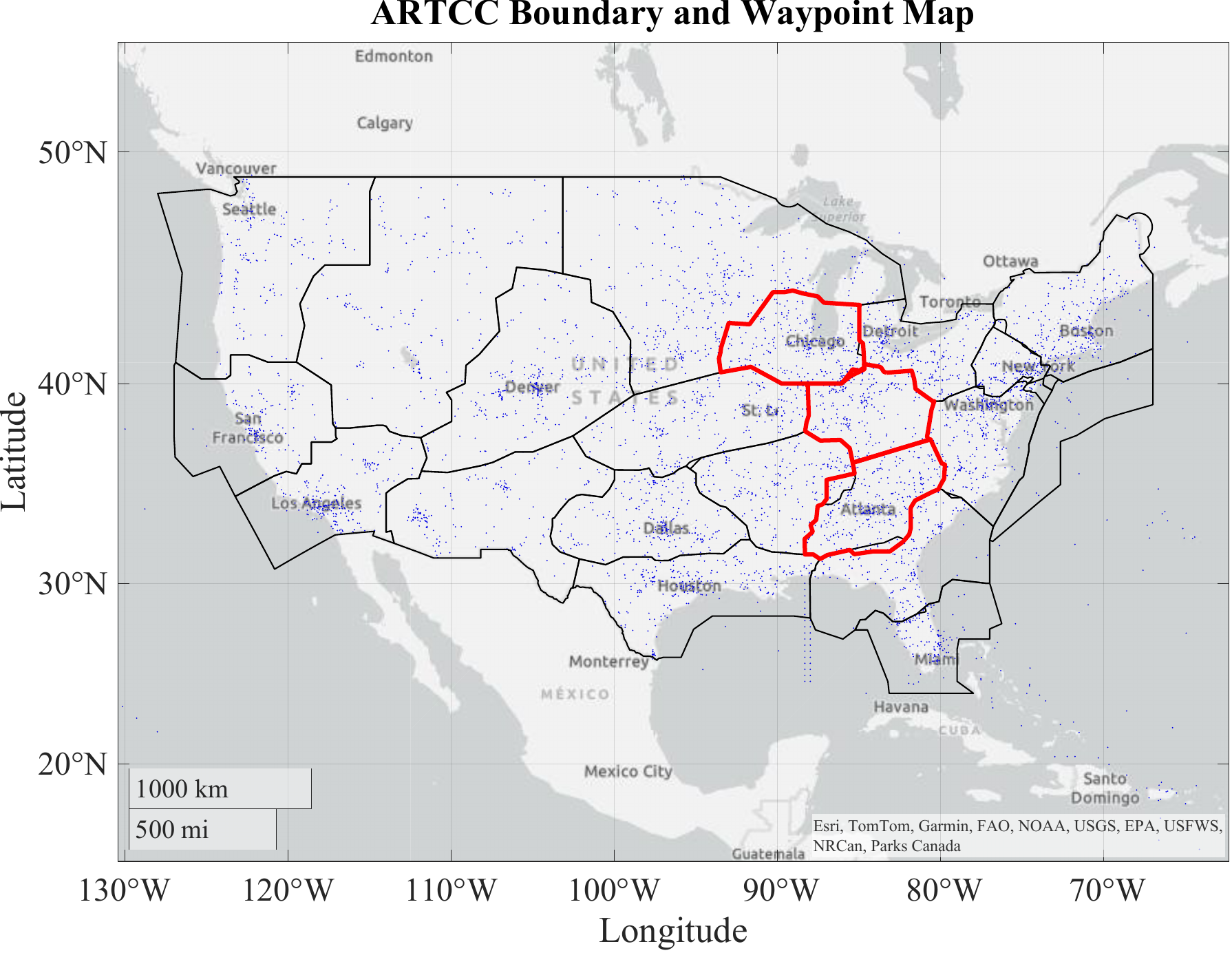}
    \caption{ARTCC boundary and waypoint map over the continental United States. 
    The map includes the boundaries of all 20 ARTCCs (black lines) 
    and 4,998 waypoints (blue dots). 
    The three ARTCCs selected for the case study—Chicago (ZAU), 
    Indianapolis (ZID), and Atlanta (ZTL)—are highlighted in red.}
    \label{fig:artcc-map}
\end{figure}

\subsection{Setup and data description} \label{sec:setup}

We consider a negotiation scenario among three neighboring ARTCCs in the United States: Chicago (ZAU), Indianapolis (ZID), and Atlanta (ZTL). These centers represent independent authorities responsible for regional airspace flow management and are modeled as self-interested agents participating in the proposed framework described in \Cref{sec:framework}. 
\Cref{fig:artcc-map} illustrates the boundaries of the 20 ARTCCs and 4,998 waypoints across the continental United States. The three ARTCCs used in this study are highlighted in red.

The agents negotiate over the TOS selections for 30 flights traversing all three ARTCCs. Each flight is associated with $q$ candidate trajectory options (here $q=5$), generated by connecting waypoints to form feasible paths across the selected ARTCC regions.

To assess the framework’s performance under varying conditions, we conduct Monte Carlo simulations with 10,000 independent trials. 
In each run, the agents’ initial asset reserves are randomly sampled from a uniform range between 1 and 20, and the taxation parameter~$\kappa$ is drawn by sampling $k \sim \mathrm{Unif}(-1,2)$ and setting $\kappa = 10^{\,k}$. 
The same ARTCC configuration and TOS options are maintained across all trials. 
The experiments evaluate  
i) how the taxation parameter~$\kappa$ and agents’ asset values~$b_i$ affect convergence rate, system efficiency, and fairness, and  
ii) how the resulting performance compares with existing centralized and decentralized coordination schemes. 

To address the large combinatorial complexity of the CTB generation, 
we employ a genetic algorithm solver implemented in Julia \cite{deep_ga}. 
\fix{The inner \texttt{TACo} parameters are fixed across all trials, with initial trading unit $d_0 = 1$, reduction factor $\gamma = 0.9$, and termination tolerance $\varepsilon = 10$.}
Experiments are conducted on a computer equipped with an Intel Core i7-12700 processor and 16~GB of RAM, with parallel computation enabled using 19 threads.

\subsection{Baselines}

\fix{We compare the proposed framework with four baselines representing centralized, operational, decentralized, and cooperative coordination philosophies within the setting introduced in \Cref{sec:setup}. 
The implementation of each baseline is described below.}

\begin{itemize}
    \item \textbf{Centralized CTOP (\texttt{C-CTOP}):} 
    A centralized optimizer determines the system-wide optimal combination of TOS assignments by minimizing the following cost function $J:\opSet \to \realSet$:
    \begin{equation} \label{eq:centCostFunction}
        J(o) = \sum_{i \in \plSet} J_i^{\mathrm{ind}}(o), \quad o\in\opSet.
    \end{equation}
    This provides a centralized performance reference under perfect information and coordination.

    \item \textbf{First-Come-First-Served CTOP (\texttt{F-CTOP}):} 
    \fix{A sequential, priority-based centralized approach that serves as an operational rationing-style baseline.} 
    Flight plans are processed in order of their scheduled departure times; the coordinator selects the most efficient trajectory option of the earlier flights first, while later flights are processed \fix{sequentially}. 
    \fix{This baseline captures the effect of rule-based priority coordination without system-wide optimization over all flights.}

    \item \textbf{Voting:} 
    A one-shot decentralized mechanism in which each ARTCC votes on the CTBs generated without oversight-guided coordination (\ie, $\coordFactor = [0,\dots,0]\in \realSet^n$). 
    The option receiving the majority of votes is selected, and ties are resolved randomly. 
    \fix{This baseline represents decentralized preference aggregation without trading, asset transfers, or regulatory correction.}

    \item \textbf{\fix{Cooperative candidate selection (\texttt{CCS}):}}
    \fix{A cooperative reference baseline that selects, from the initial CTB candidate pool generated without oversight-guided coordination, the CTB with the minimum aggregate system cost. 
    This baseline uses the same locally generated candidates as the Voting baseline, but replaces decentralized preference aggregation or \texttt{TACo} negotiation with centralized cooperative selection.}
\end{itemize}

\fix{\Cref{tab:baseline_summary} summarizes the decision rule and comparison role of each baseline.}

\begin{table}[hbt!]
\centering
\caption{\fix{Summary of baseline roles in the CTOP case study.}}
\label{tab:baseline_summary}
\begin{tabular}{lll}
\hline
\fix{\textbf{Method}} & \fix{\textbf{Decision rule}} & \fix{\textbf{Role in comparison}}
\\
\hline
\fix{\texttt{C-CTOP}}
& \fix{Minimize aggregate cost over $\opSet$} 
& \fix{Centralized system-optimal reference} \\

\fix{\texttt{F-CTOP}} 
& \fix{Sequential priority-based assignment} 
& \fix{Operational rationing-style reference} \\

\fix{Voting} 
& \fix{Majority vote over initial CTBs} 
& \fix{Decentralized aggregation without trading} \\

\fix{\texttt{CCS}}
& \fix{Minimize aggregate cost over initial CTBs} 
& \fix{Cooperative selection without negotiation} \\
\hline
\end{tabular}
\end{table}

\subsection{Results}
We first examine how the taxation parameter and agents’ asset valuations shape the behavior and outcomes of the proposed framework.
\subsubsection{Effect of taxation $\kappa$ and asset valuation $b_i$}
\begin{figure}[hbt!]
    \centering
    \includegraphics[width=0.98\linewidth]{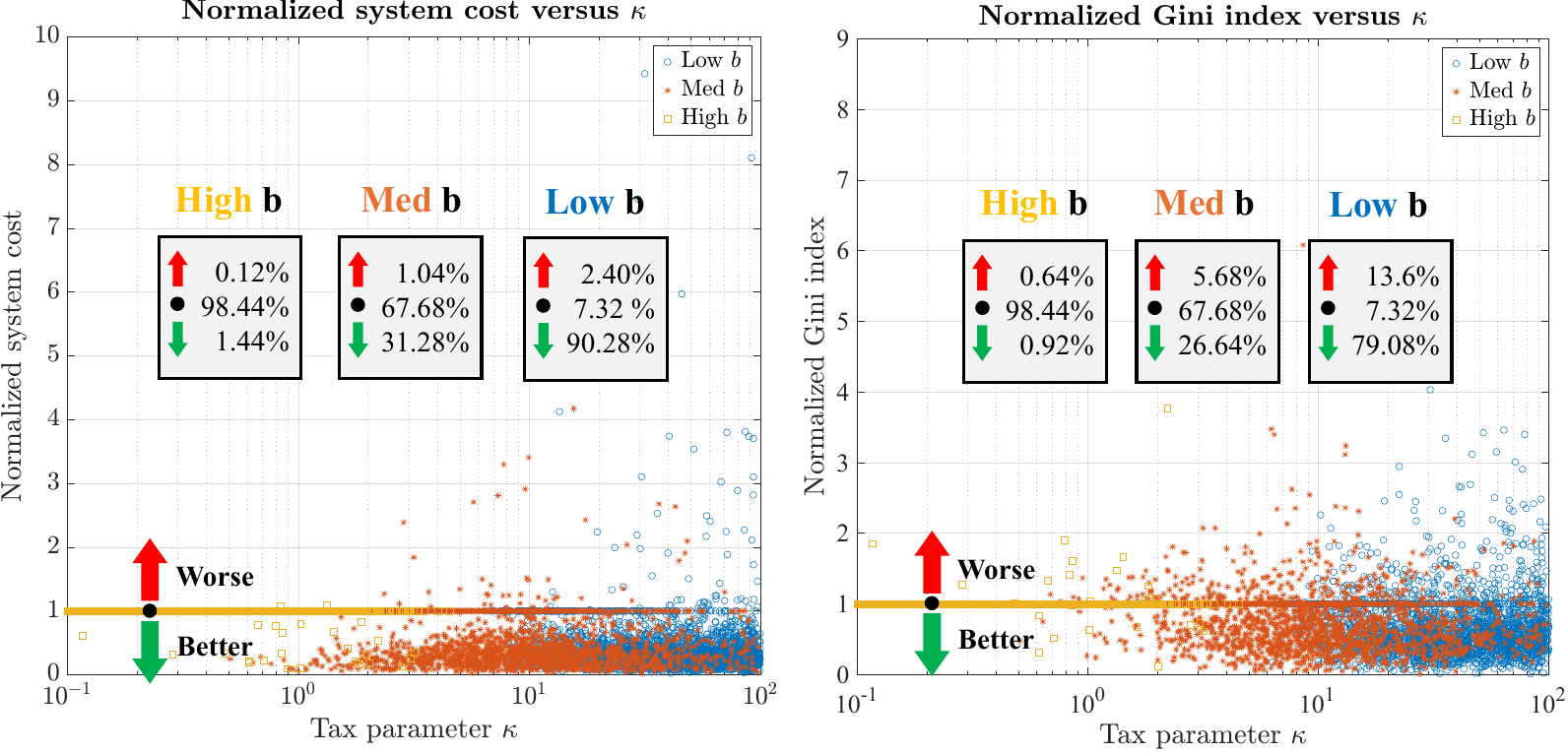}
    \caption{
    Relative system cost and fairness improvement across $\kappa$ values. 
    Each point represents the final value normalized by the performance after first-iteration, 
    grouped by the agents’ average asset value $\bar b = \sum_{i\in\plSet}{b_i}/n$. 
    The colors indicate the asset value groups: 
    blue for low $b$, orange for medium $b$, 
    and yellow for high $b$. 
    Larger $\kappa$ values lead to slower but more balanced convergence, 
    reducing both system cost and Gini index. 
    Boxed numbers indicate the proportion of cases that improved ($<1$), 
    remained unchanged ($=1$), or degraded ($>1$).
}
    \label{fig:result-internal}
\end{figure}

\Cref{fig:result-internal} illustrates how the taxation parameter~$\kappa$ and the agents’ asset values~$b_i$ influence convergence behavior, system efficiency, and fairness. Each data point corresponds to one simulation, showing the ratio of the performance metric of the final outcome ($o^{\texttt{TACo}}_{r_{term}}$) to that achieved in the first iteration ($o^{\texttt{TACo}}_{1}$). Hence, a value of~$1$ indicates immediate convergence, while a value below~$1$ indicates system improvement through additional negotiation rounds.

Across simulations, we observe relationships between the taxation parameter~$\kappa$, the agents’ asset values~$b_i$, convergence rate, and the system-level performance. 
As~$\kappa$ increases, the average asset value~$b_i$ naturally decreases, reflecting the inverse relationship $b_i = 1/(\kappa R_i)$. 
When $b$ is large (smaller $\kappa$), most cases ($98.44\%$) converge within a single round, whereas smaller~$b$ (induced by larger $\kappa$) values lead to multiple negotiation rounds, in line with \Cref{eq:round-upper}. 
\fix{This low-$\kappa$ regime also serves as a near-\texttt{TACo}-only reference: since $98.44\%$ of these cases terminate after the first iteration, the outcome is usually determined by the initial \texttt{TACo} negotiation before any coordination-factor update takes effect.}

With additional iterations, a substantial portion of trials achieve improvement in both system efficiency and fairness: in the low-$b$ group (which arises from larger $\kappa$ or reserves, or both), $90.28\%$ of cases yield lower final system costs and $79.08\%$ achieve reduced Gini index values (smaller Gini values indicate more equitable outcomes \cite{fairnessGuide}) compared to the first iteration.
Although a small number of runs exhibit degraded performance, such cases remain relatively rare across all~$\kappa$ values. 
These results support that~$\kappa$ regulates both the convergence rate and the extent of achievable performance enhancement, consistent with the theoretical results in \Cref{corollary:round-upper} and \Cref{lem:sys-gap}.

\subsubsection{Baseline comparison}

\begin{figure}[hbt!]
    \centering
    \begin{subfigure}{0.475\textwidth}
        \includegraphics[width=\linewidth, trim=0 0 -50 0, clip]{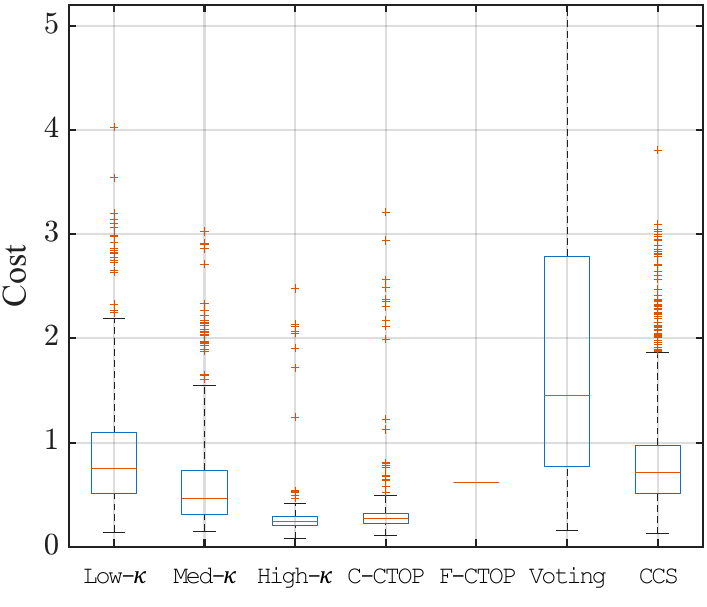}
        \caption{System cost comparison across $\kappa$ values and baselines.}
        \label{fig:baseline-cost}
    \end{subfigure}
    \hfill
    \begin{subfigure}{0.49\textwidth}
        \includegraphics[width=\linewidth, trim=0 0 -50 0, clip]{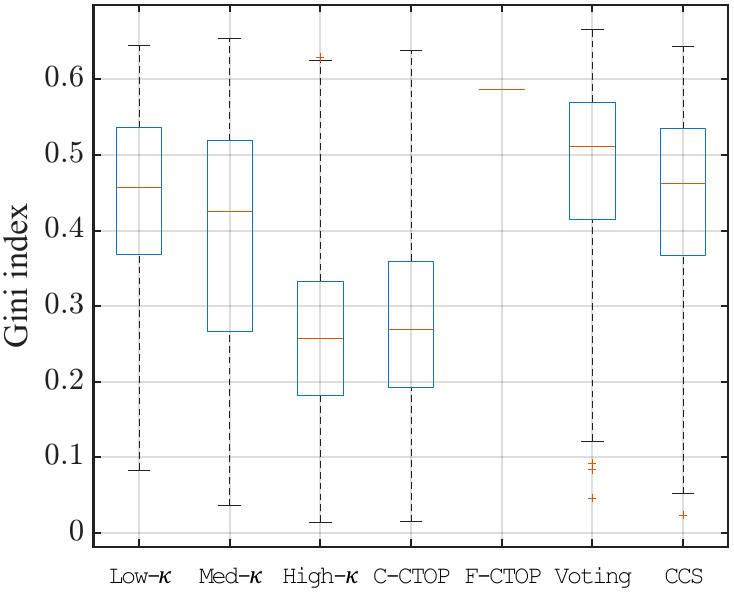}
        \caption{Fairness (Gini index) comparison.}
        \label{fig:baseline-fairness}
    \end{subfigure}
    \caption{
    Performance comparison between the proposed decentralized oversight framework and baseline coordination mechanisms. 
    \fix{The low-, medium-, and high-$\kappa$ groups show the effect of increasing oversight strength, while \texttt{C-CTOP}, \texttt{F-CTOP}, Voting, and \texttt{CCS} provide centralized, operational, decentralized, and cooperative references.}}
    \label{fig:result-baseline}
\end{figure}

\Cref{fig:result-baseline} compares the performance of the proposed framework with \fix{the four representative baselines summarized in \Cref{tab:baseline_summary}: \texttt{C-CTOP}, \texttt{F-CTOP}, Voting, and \texttt{CCS}.}
The comparison is conducted under three $\kappa$ regimes, defined by the quartiles of the tax parameter $\kappa$: the lowest quartile (Q1), the interquartile range (Q2--Q3), and the highest quartile (Q4).

\fix{Consistent with the near-\texttt{TACo}-only interpretation in \Cref{fig:result-internal}, the low-$\kappa$ regime mostly reflects first-round \texttt{TACo} behavior, while larger $\kappa$ values enable additional oversight-guided candidate generation.}
As $\kappa$ increases, the proposed framework achieves lower system cost and smaller Gini index, closely approaching the centralized reference solution at high~$\kappa$.
Interestingly, in some runs, the decentralized oversight algorithm slightly outperforms the centralized solver due to the stochastic exploration in the genetic algorithm-based choice generation process.  
Note that this result does not imply that the centralized approach is inferior, but rather that the proposed approach can attain near-optimal performance while preserving agent autonomy and \fix{avoiding direct disclosure of private asset valuations}.

\fix{\texttt{CCS} exhibits performance comparable to the low-$\kappa$ regime, suggesting that first-round \texttt{TACo}-dominant behavior already selects outcomes similar to cooperative selection over the same initial CTB pool. 
However, the higher-$\kappa$ regimes further reduce both system cost and Gini index, indicating that the main benefit of the oversight loop comes from multi-round, oversight-guided candidate regeneration rather than from centralized selection over the initial candidates alone.}

In contrast, both \fix{\texttt{F-CTOP}} and Voting exhibit weaker performance.
The \fix{\texttt{F-CTOP}} policy achieves system costs comparable to the low-$\kappa$ regime, illustrating that purely sequential coordination sacrifices global optimality.
The Voting baseline performs worse: it yields substantially higher system cost with large variance, and its Gini index remains elevated and often comparable to that of \texttt{F-CTOP}, indicating less favorable fairness as well.
\fix{Overall, these comparisons show that the proposed framework occupies a middle ground between centralized system optimization, cooperative selection over initially generated candidates, and purely decentralized aggregation, while the improvement at higher $\kappa$ highlights the value of iterative oversight-guided candidate generation.}

\begin{figure}[hbt!]
    \centering
    \begin{subfigure}{0.48\textwidth}
        \includegraphics[width=\linewidth]{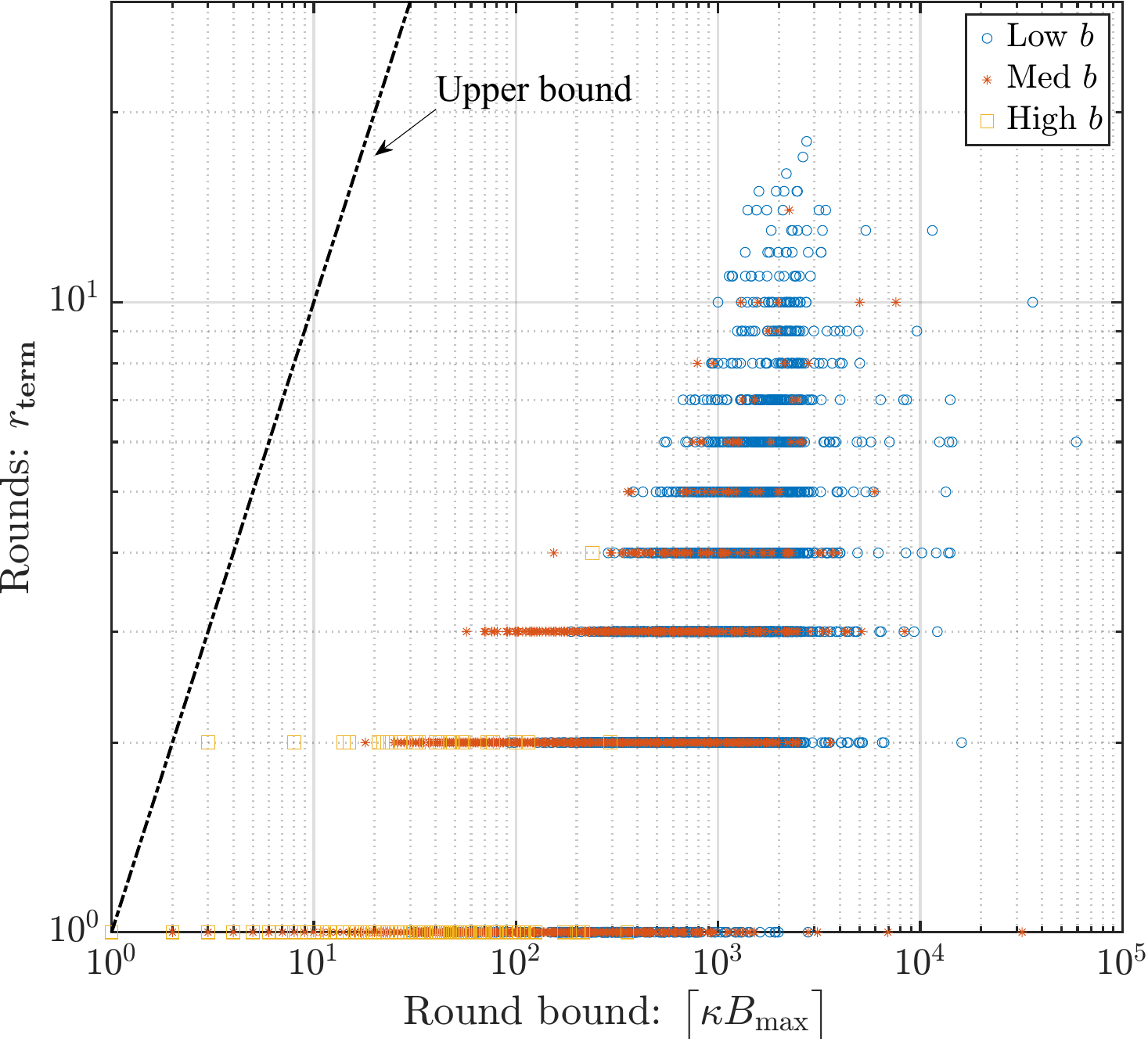}
        \caption{Round-bound verification.}
        \label{fig:round-bound-plot}
    \end{subfigure}
    \hfill
    \begin{subfigure}{0.48\textwidth}
        \includegraphics[width=\linewidth]{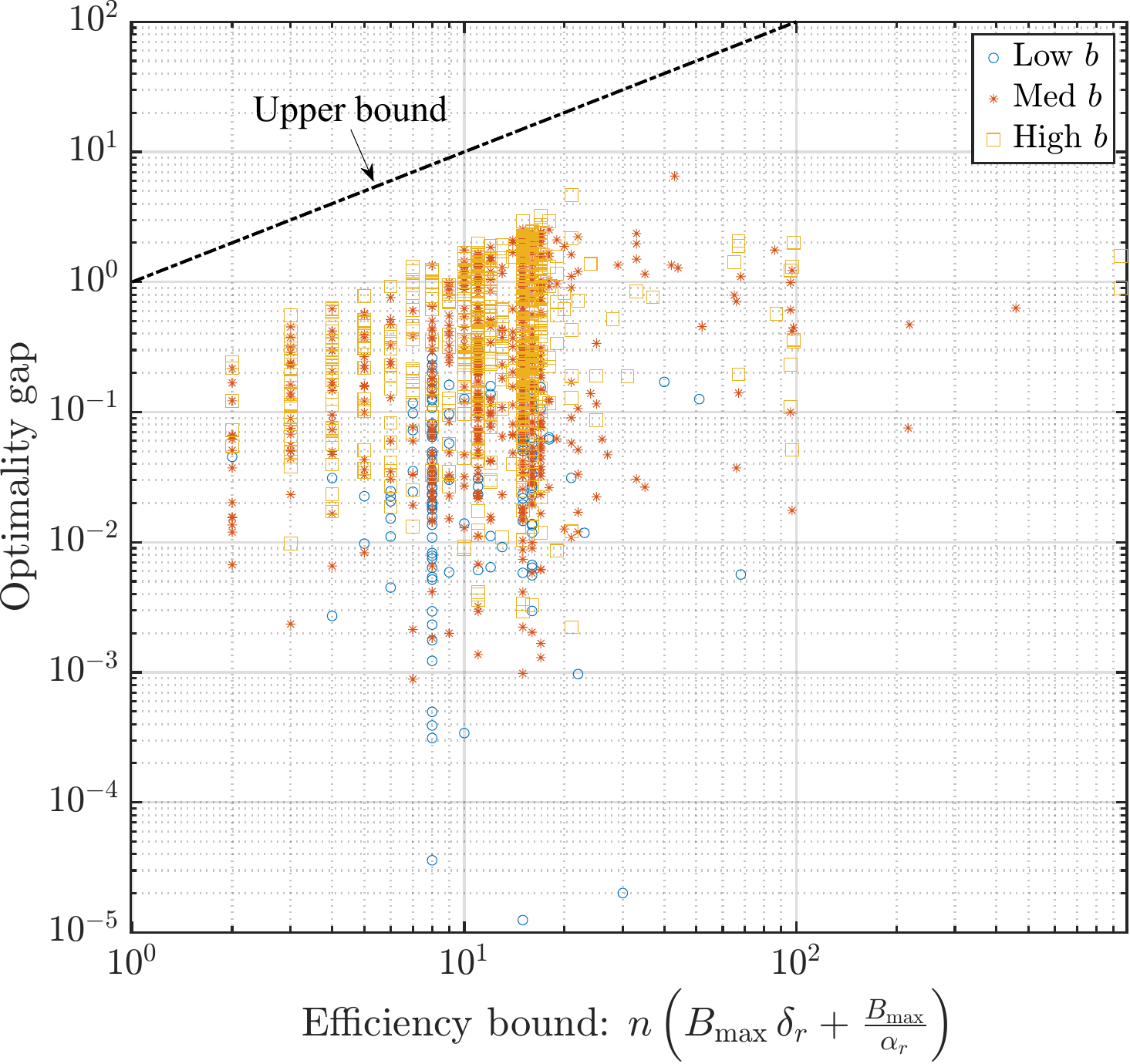}
        \caption{Optimality-gap bound verification.}
        \label{fig:gap-bound-plot}
    \end{subfigure}
    \caption{
    Validation of theoretical bounds derived in \Cref{eq:round-upper,eq:gap-bound-by-round}. 
    Each point represents a simulation instance, where the horizontal axis shows the analytical bound 
    and the vertical axis shows the observed value. 
    All samples remain below theoretical upper bounds, confirming the theoretical results in \Cref{subsec:kappa-rate} and \Cref{subsec:kappa-eff}.}
    \label{fig:result-bound}
\end{figure}

\subsubsection{Bound verification}

Finally, \Cref{fig:result-bound} validates the analytical bounds derived in \Cref{subsec:kappa-rate,subsec:kappa-eff}.  
The scatter plots compare the theoretical upper bounds (horizontal axis) with the observed outcomes (vertical axis) for (a) the number of negotiation rounds and (b) the optimality gap. All data points remain below the upper bound line, confirming that the theoretical limits hold in practice.

\subsubsection{\fix{Computation time analysis}}
\label{subsec:runtime}

\fix{We report wall-clock computation times by framework stage for the CTOP case study in \Cref{tab:runtime}. The timing is decomposed into candidate generation, \texttt{TACo} negotiation, and shortfall-based update.}

\begin{table}[hbt!]
\centering
\caption{\fix{Computation time by framework stage in the CTOP case study. Candidate-generation times are reported as per-agent averages over the three ARTCC agents. Times are reported as wall-clock time; P95 denotes the 95th percentile.}}
\label{tab:runtime}
\fix{
\begin{tabular}{lrrrr}
\toprule
Component & Mean [ms] & Median [ms] & P95 [ms] & Share [\%] \\
\midrule
Candidate generation & 12646.61 & 4001.19 & 48639.22 & >\,99.99 \\
\texttt{TACo} negotiation & 0.76 & 0.03 & 3.27 & <\,0.01 \\
Shortfall/update & 0.00 & 0.00 & 0.01 & <\,0.01 \\
Total & 12647.37 & 4001.22 & 48642.50 & -- \\
\bottomrule
\end{tabular}
}
\end{table}

\fix{Since candidate generation is performed independently by each ARTCC agent, Table~\ref{tab:runtime} reports candidate-generation time as the per-agent average over the three agents. 
In the scenario considered in this experiment, the median total computation time is approximately $4.0$ seconds, with a 95th percentile of approximately $48.6$ seconds. 
The runtime is dominated by CTB generation: the median candidate-generation time is approximately $4.0$ seconds, whereas the median \texttt{TACo} negotiation time is $0.03$ ms and the median shortfall/update time is below the millisecond scale.
Thus, in this implementation, the proposed negotiation and oversight stages add negligible computational overhead relative to CTB generation.}

\fix{The CTB-generation module serves as a proxy candidate generator for the case study, as described in \Cref{sec:setup}. 
Therefore, the total runtime can be reduced by replacing the proxy with specialized trajectory-bundle generation methods. 
This runtime analysis should be interpreted as computational runtime only; it does not include infrastructure-related delays, such as communication latency between ARTCCs. 
Overall, these results suggest that, at the scale considered in this case study, the computational component of the framework fits within a minute-level tactical decision-support timescale.}

\section{Conclusion}
\label{sec:conclusion}
We propose an iterative negotiation and oversight framework that enables coordination among self-interested agents while safeguarding system-level performance through taxation-like intervention. 
We establish theoretical guarantees of finite-round convergence and derive explicit bounds linking the level of central intervention to both convergence rate and system efficiency. 
These results provide practical guidance for a central coordinator in selecting the intervention level to balance convergence speed, efficiency, and fairness.

To validate the framework, we developed a decentralized variant of the \textit{collaborative trajectory options program} (CTOP)—a traffic management initiative currently implemented in the United States national airspace system—and evaluated it in a realistic sector-level coordination scenario. 
Simulation results confirm that the framework converges in all cases and that adjusting the intervention level effectively regulates the convergence rate, efficiency, and fairness. 
The oversight does not directly optimize the system but implicitly guides agents toward coordinated behavior, thereby \fix{preserving decentralized negotiation while correcting system-level deviations through limited regulatory guidance}. 
Overall, the results demonstrate a practical pathway for decentralized coordination in air traffic management.

The same principle can also be generalized beyond air traffic management to other multi-agent systems that require reliable \fix{coordination among self-interested agents under limited oversight}. 
Future work may focus on tightening the theoretical bounds, extending the framework to multi-asset trading schemes, and developing decentralized safeguard mechanisms that ensure fairness and efficiency with reduced reliance on centralized oversight or taxation. 
\fix{Additional directions include dynamic reserve design, robustness against malicious or non-compliant agents, event-triggered or rolling-horizon operation under changing weather and capacity conditions, and faster candidate-generation methods for large-scale deployment.}

\section*{CRediT authorship contribution statement}
\noindent
\textbf{Jaehan Im}: Conceptualization, Methodology, Software, Formal analysis, Writing – original draft. 
\textbf{John-Paul Clarke}: Supervision, Methodology, Writing – review \& editing. 
\textbf{Ufuk Topcu}: Supervision, Funding acquisition, Methodology, Writing – review \& editing.
\textbf{David Fridovich-Keil}: Supervision, Funding acquisition, Methodology, Writing – review \& editing. 

\section*{Declaration of competing interest}
The authors declare that they have no known competing financial interests or personal relationships that could have appeared to influence the work reported in this paper.

\section*{Acknowledgments}
\noindent
This work was supported by the National Aeronautics and Space Administration ULI Award under grant 80NSSC21M0071 and 80NSSC24M0070, and by the National Science Foundation CAREER award under grants 2336840 and 2211548.

\appendix
\section{Proofs for the Theorems and Lemmas} \label{app1}

\begin{proof}[\textbf{Proof of \Cref{lemma:choice-conv}}]
Let $\opSet$ denote the finite set of all feasible choices, and let 
$\mathcal{O}^{(r)} \subseteq \opSet$ denote the set of choices selected by at least one agent at round~$r$. Recall that at outer iteration $r$ the effective cost for agent $i$ is
\begin{equation}
J_i^{w^{(r)}}(o)
    = \frac{1}{\alpha_r}
    \Big(J_i^{\mathrm{ind}}(o) 
        + S_{w^{(r)}}(o)\Big),
\quad 
S_{w^{(r)}}(o) := (w^{(r)})^\top J^{\mathrm{ind}}(o),
\end{equation}
where $\alpha_r = \|w^{(r)}\|_1+1$ and $\alpha_{r+1}=\alpha_r+1$.

Since $w^{(r)} \ge 0$, we have
\begin{equation}
|S_{w^{(r)}}(o)-S_{w^{(r)}}(o')|
    \le \|w^{(r)}\|_1 
        \, \max_{i} |J_i^{\mathrm{ind}}(o)-J_i^{\mathrm{ind}}(o')|
    \le \alpha_r B_{\max},
\end{equation}
where 
\begin{equation}
B_{\max}
    := \max_{i} \max_{o,o'\in\opSet} 
       |J_i^{\mathrm{ind}}(o)-J_i^{\mathrm{ind}}(o')|
    < \infty,
\end{equation}
using boundedness of intrinsic costs.

Hence, for any $o,o'\in\mathcal{O}^{(r)}$,
\begin{equation}
\big| J_i^{w^{(r)}}(o)-J_i^{w^{(r)}}(o') \big|
    \le 
    \frac{1}{\alpha_r}
    \Big(
        |J_i^{\mathrm{ind}}(o)-J_i^{\mathrm{ind}}(o')|
        + |S_{w^{(r)}}(o)-S_{w^{(r)}}(o')|
    \Big)
    \le \frac{B_{\max}}{\alpha_r}.
\end{equation}

Thus,
\begin{equation}
\max_{o,o'\in\mathcal{O}^{(r)}}
    |J_i^{w^{(r)}}(o)-J_i^{w^{(r)}}(o')|
        \xrightarrow[r\to\infty]{} 0,
\end{equation}
since $\alpha_r \to \infty$.  
This shows that the effective cost spread within $\mathcal{O}^{(r)}$ shrinks to zero as $r$ increases.
\end{proof}

\begin{proof}[\textbf{Proof of \Cref{lemma:ind-transfer}}]
Fix round $r$ and let $t_i^{(r)}$ be the net transfer received by agent $i$ under
the \texttt{TACo} outcome $o_r^{\texttt{TACo}}$.
For each agent~$i$, let
\begin{equation}\label{eq:best-alt}
o_i^{(1)} 
\in 
\arg\min_{o\in\opSet\setminus\{o_r^{\texttt{TACo}}\}}
J_i^{w^{(r)}}(o)
\end{equation}
denote its best alternative, and define the corresponding gap
\begin{equation}
\Delta_i^{(r)}
:=
J_i^{w^{(r)}}(o_i^{(1)})
-
J_i^{w^{(r)}}(o_r^{\texttt{TACo}}).
\end{equation}

\texttt{TACo} satisfies \emph{individual rationality}, i.e.,
\begin{equation}\label{eq:IR-short}
J_i^{w^{(r)}}(o_r^{\texttt{TACo}})-t_i^{(r)}
\le 
J_i^{w^{(r)}}(o_i^{(1)}),
\end{equation}
which is equivalent to
\begin{equation}
-t_i^{(r)} \le \Delta_i^{(r)}.
\end{equation}
Thus, if $i$ is a beneficiary ($t_i^{(r)}>0$), then 
\begin{equation}
t_i^{(r)} \ge \ell_i^{(r)} := \max\{0,-\Delta_i^{(r)}\},
\end{equation}
and if $j$ is a payer ($t_j^{(r)}<0$), then
\begin{equation}
|t_j^{(r)}|
\le 
\fix{v_j}^{(r)} := \max\{0,\Delta_j^{(r)}\}.
\end{equation}
This yields the claimed lower and upper transfer bounds for beneficiaries and payers.
\end{proof}

\begin{proof}[\textbf{Proof of \Cref{cor:round-bound}}]
From \Cref{lemma:ind-transfer}, the magnitude of the transfer paid by any
payer $j\in W_r$ is bounded by
\begin{equation}
    |t_j^{(r)}|
    \le
    \fix{v_j}^{(r)}
    :=
    \max\{0,\Delta_j^{(r)}\},
\end{equation}
where $\Delta_j^{(r)}$ is the difference between agent $j$’s
best alternative cost and its cost under the \texttt{TACo} outcome.

By \Cref{lemma:choice-conv}, the effective cost spread at round $r$ satisfies
\begin{equation}
    |\Delta_j^{(r)}|
    \le
    \max_{o,o'\in\opSet}
    \big|J_j^{w^{(r)}}(o)-J_j^{w^{(r)}}(o')\big|
    \le
    \frac{B_{\max}}{\alpha_r}.
\end{equation}
Thus every payer obeys
\begin{equation}
    0 \le |t_j^{(r)}|
    \le \fix{v_j}^{(r)}
    \le \frac{B_{\max}}{\alpha_r},
    \qquad j\in W_r,
\end{equation}
establishing the desired round-wise transfer bound.
\end{proof}

\begin{proof}[\textbf{Proof of \Cref{theorem:termination-ind}}]
By \Cref{cor:round-bound}, every payer $j \in W_r$ satisfies
\begin{equation}
    |t_j^{(r)}| \;\le\; \frac{B_{\max}}{\alpha_r}.
\end{equation}
Transfers are quantized as 
\begin{equation}
    |t_j^{(r)}| = c_j^{(r)} b_j,
    \qquad b_j = \frac{1}{\kappa R_j},
\end{equation}
so
\begin{equation}
    c_j^{(r)} \;\le\; \frac{B_{\max}}{\alpha_r}\,\kappa R_j.
\end{equation}

Since $\alpha_r$ increases by one each round, there exists $\bar r$ such that
$\alpha_r \ge \kappa B_{\max}$ for all $r \ge \bar r$.  
For these rounds,
\begin{equation}
    c_j^{(r)} \;\le\; R_j,
\end{equation}
so no payer exceeds its reserve constraint.  
Thus, the process must terminate in finite rounds.
\end{proof}

\begin{proof}[\textbf{Proof of \Cref{theorem:tax-to-alpha}}]
From the proof of \Cref{theorem:termination-ind}, we have for any payer $j$
and round $r$ that
\begin{equation}
    c_j^{(r)} \;\le\; \frac{B_{\max}}{\alpha_r}\,\kappa R_j.
\end{equation}
Therefore, whenever $\alpha_r \ge \kappa B_{\max}$, it follows that
\begin{equation}
    c_j^{(r)} \;\le\; R_j
\end{equation}
for all payers $j$, so every reserve constraint is satisfied and the process
terminates. 
\end{proof}

\begin{proof}[\textbf{Proof of \Cref{corollary:round-upper}}]
Since $\alpha_1 = 1$ and $\alpha_r$ increases by one at each round, we have $\alpha_r = r$.
By \Cref{theorem:tax-to-alpha}, the process terminates once $\alpha_r \ge \kappa B_{\max}$,
that is, once $r \ge \kappa B_{\max}$.
Hence the termination round satisfies
\begin{equation}
    r_{\mathrm{term}} \;\le\; \big\lceil \kappa B_{\max} \big\rceil.
\end{equation}
Equivalently, for a desired iteration budget $r_{\mathrm{des}}$, termination is guaranteed if
\begin{equation}
    \kappa \;\le\; \frac{r_{\mathrm{des}}}{B_{\max}}.
\end{equation}
\end{proof}

\begin{proof}[\textbf{Proof of \Cref{lemma:eta-bound}}]
Let $o^\dagger \in \arg\min_{o} S_{\bar{w}^{(r)}}(o)$ and let $i^\star$ be the supporting
player for the \texttt{TACo} outcome $o_r^{\texttt{TACo}}$, i.e.,
\begin{equation}
S_{\bar{w}^{(r)}}(o_r^{\texttt{TACo}})
    + \frac{1}{\alpha_r} J_{i^\star}^{\mathrm{ind}}(o_r^{\texttt{TACo}})
\;\le\;
S_{\bar{w}^{(r)}}(o^\dagger)
    + \frac{1}{\alpha_r} J_{i^\star}^{\mathrm{ind}}(o^\dagger).
\end{equation}
Rearranging yields
\begin{equation}
0 \;\le\;
S_{\bar{w}^{(r)}}(o_r^{\texttt{TACo}}) - S_{\bar{w}^{(r)}}(o^\dagger)
\;\le\;
\frac{
    J_{i^\star}^{\mathrm{ind}}(o^\dagger)
    - J_{i^\star}^{\mathrm{ind}}(o_r^{\texttt{TACo}})
}{\alpha_r}.
\end{equation}
Since intrinsic costs differ by at most $B_{\max}$ over $\opSet$,
\begin{equation}
0 \le 
S_{\bar{w}^{(r)}}(o_r^{\texttt{TACo}})
    - S_{\bar{w}^{(r)}}(o^\dagger)
\;\le\; \frac{B_{\max}}{\alpha_r}.
\end{equation}
Thus the selection error satisfies $\eta_r \le B_{\max}/\alpha_r$.
\end{proof}

\begin{proof}[\textbf{Proof of \Cref{lem:sys-gap}}]
Recall $J_{\mathrm{sys}}(o)=n S_u(o)$ where $u=\tfrac{1}{n}\mathbf{1}$.
For any weight vector $\bar w^{(r)} = w^{(r)}/\alpha_r$, we may write
\begin{equation}
    S_u(o_r^{\texttt{TACo}}) - S_u(o^{\mathrm{opt}})
    =
    \big(\bar w^{(r)} - u\big)^\top
        \big(J^{\mathrm{ind}}(o_r^{\texttt{TACo}})
            - J^{\mathrm{ind}}(o^{\mathrm{opt}})\big)
    + \eta_r,
\end{equation}
where $\eta_r = S_{\bar w^{(r)}}(o_r^{\texttt{TACo}}) -
                 S_{\bar w^{(r)}}(o^\dagger)$
and $o^\dagger \in \arg\min_o S_{\bar w^{(r)}}(o)$.
Taking absolute values and using
$\|J^{\mathrm{ind}}(o_r^{\texttt{TACo}}) -
     J^{\mathrm{ind}}(o^{\mathrm{opt}})\|_\infty \le B_{\max}$ and
$\delta_r=\|\bar w^{(r)}-u\|_1$ gives
\begin{equation}
0 \le
S_u(o_r^{\texttt{TACo}}) - S_u(o^{\mathrm{opt}})
\le
B_{\max}\,\delta_r + \eta_r.
\end{equation}
Multiplying by $n$ yields
\begin{equation}
0 \le
J_{\mathrm{sys}}(o_r^{\texttt{TACo}})
    - J_{\mathrm{sys}}(o^{\mathrm{opt}})
\le
n\big(B_{\max}\,\delta_r + \eta_r\big),
\end{equation}
which proves the claim.
\end{proof}

\begin{proof}[\textbf{Proof of \Cref{thm:necessary-kappa}}]
Since $\alpha_r = r$ and $r_{\max} = \lceil \kappa B_{\max} \rceil$, 
assume $\delta_r \le \bar{\delta}$ for all $r \le r_{\max}$.
From \Cref{lem:sys-gap},
\begin{equation} \label{eq:sys-gap-necessary}
    J_{\mathrm{sys}}(o_r^{\texttt{TACo}})
        - J_{\mathrm{sys}}(o^{\mathrm{opt}})
    \;\le\;
    n\!\left(B_{\max}\bar{\delta} + \frac{B_{\max}}{r}\right).
\end{equation}
To achieve a target system gap $\tau$, it is necessary that 
$\tau > nB_{\max}\bar{\delta}$ and
\begin{equation}
    r \;\ge\; \frac{nB_{\max}}{\tau - nB_{\max}\bar{\delta}}.
\end{equation}
Because $r_{\mathrm{term}} \le r_{\max} = \lceil \kappa B_{\max}\rceil$, 
the above condition requires
\begin{equation}
    \lceil \kappa B_{\max}\rceil 
    \;\ge\; 
    \frac{nB_{\max}}{\tau - nB_{\max}\bar{\delta}},
\end{equation}
which yields the stated lower bound on $\kappa$.
\end{proof}

\begin{proof}[\textbf{Proof of \Cref{cor:kappa-geometry}}]
Since $u=\tfrac{1}{n}\mathbf{1}$ and $\bar w^{(r)} \in \Delta^n$, the weight
misalignment satisfies
\begin{equation}
    \delta_r = \|\bar w^{(r)} - u\|_1 \;\le\; 2\!\left(1 - \tfrac{1}{n}\right)
    =: \sigma_{\max}.
\end{equation}
Substituting $\bar{\delta} = \sigma_{\max}$ into the bound of
\Cref{thm:necessary-kappa} (valid when $\tau > 2(n-1)B_{\max}$) gives
\begin{equation}
    \kappa 
    \;\ge\;
    \frac{1}{B_{\max}}
    \left\lfloor
        \frac{nB_{\max}}
             {\tau - 2(n-1)B_{\max}}
    \right\rfloor,
\end{equation}
which yields the stated geometric lower bound on $\kappa$.
\end{proof}

{\small
\bibliographystyle{elsarticle-num} 
\bibliography{reference}
}



\end{document}